\documentclass[aps,pre,amsfonts,amssymb,amsmath,twocolumn, superscriptaddress]{revtex4-1}
\usepackage{color}
\usepackage{environ}
\usepackage{graphicx}
\usepackage{hyperref}
\usepackage{MnSymbol}
\usepackage{multirow}
\usepackage{supertabular}

\graphicspath{{}{../figures/}{figures/}}
\newcommand{\al}[1]{\begin{align} #1 \end{align}} 
\newcommand{\alnn}[1]{\begin{align*} #1 \end{align*}} 
\newcommand{\algrp}[2]{\begin{subequations}\eqlabel{#1}\begin{align}#2\end{align}\end{subequations}}
\newcommand{\aleq}[2]{\begin{equation}\eqlabel{#1}\begin{split}#2\end{split}\end{equation}}

\newcommand{\ie}{i.e.}
\newcommand{\eg}{e.g.}

\let\ring\r
\newcommand{\eqlabel}[1]{\label{eq:#1}}
\newcommand{\eq}[1]{\eqref{eq:#1}}
\NewEnviron{eqsplit}[1]{\begin{subequations}\eqlabel{#1}\begin{align} \BODY \end{align}\end{subequations}}
\newcommand{\nn}{\nonumber}
\newcommand{\millim}{\ensuremath{\mathrm{mm}}} % millimeter
\newcommand{\avg}[1]{\overline{#1}}     % average over psi, result independent of psi
\newcommand{\conj}[1]{\overline{#1}}    % complex conjugate TODO: identical to avg? 
\newcommand{\slavg}[1]{\underline{#1}}  % sliding average over one period
\newcommand{\cen}[1]{\overset\circ{#1}} % of centre frame
\newcommand{\tip}[1]{\check{#1}}        % of tip frame
\newcommand{\base}[1]{#1_*}             % unperturbed 
\newcommand{\pert}[1]{\tilde{#1}}       % perturbation
\newcommand{\trans}[1]{\tilde{#1}}      % transformed by group element

\newcommand{\mean}[1]{\langle #1 \rangle}       % average over positions of meander pattern
\newcommand{\bbraket}[2]{\llangle#1\mid#2\rrangle}  % long inner product
\newcommand{\braket}[2]{\langle #1\mid #2\rangle}               % short inner product 
\providecommand{\abs}[1]{\left\lvert #1 \right\rvert}
\newcommand{\bzero}{\boldsymbol{0}}     % zero column
\newcommand{\bone}{\boldsymbol{1}}     % identity matrix
\newcommand{\const}{\mathrm{const}} % constant
\renewcommand{\d}{\mathrm{d}}           % ordinary differential
\newcommand{\dd}{\partial}              % partial differential 
\newcommand{\diag}{\mathrm{diag}}       % diagonal mx
\newcommand{\iu}{i}                      % imaginary unit
\newcommand{\sgn}{\mathop{\mathrm{sgn}}} % signum function
\newcommand{\Lap}{\Delta}               % Laplacian
\newcommand{\Real}{\mathbb{R}}          % set of reals
\newcommand{\OO}{\mathcal{O}}           % asymptotic order
\newcommand{\Tr}{\mathrm{Tr}}           % trace of a matrix
\newcommand{\Zahlen}{\mathbb{Z}}        % set of integers
\newcommand{\df}[2]{\frac{\partial #1}{\partial #2}}
\newcommand{\e}{e}                  % nat log base
\newcommand{\Mx}[1]{\left(\begin{array}{ccccccc}#1\end{array}\right)}
\ifx\notcolor\undefined\newcommand{\notcolor}{blue}\else\fi
\newcommand{\+}[2]{\def#1{{\color{\notcolor}#2}}}
\newcommand{\1}[2]{\def#1##1{{\color{\notcolor}#2}}}
\newcommand{\2}[2]{\def#1##1##2{{\color{\notcolor}#2}}}

\+{\A}{A}\+{\B}{B}\+{\C}{C}\+{\D}{D} % corot frame Cartesian coord indices
\+{\Aconst}{A}                      % a constant in locking condition 
\+{\aconst}{\alpha}                 % another constant inlocking condition
\+{\a}{a} \+{\b}{b}                 % lab frame Cartesian coord indices
\+{\adj}{^\dagger}                  % operator adjoint
\+{\alp}{\beta}                     % meandering angle
\+{\alpo}{\alp_0}                   % .., geometric (minimal angle)
\+{\alps}{\alp_s}                   % .., kinematic (spiral turning angle)
\+{\apar}{a}\+{\bpar}{b}\+{\cpar}{\varsigma} % Barkley parameters, c=epsilon
\+{\fhneps}{\alpha}\+{\fhnbet}{\beta}\+{\fhngam}{\gamma} % FHN parameters, alpha=epsilon
\+{\AGroup}{B}                      % a group
\+{\bF}{\mathbf{F}}                 % reaction rates column-function
\+{\bQ}{\mathbf{P}}                 % "any 2x2 matrix" in Sec VIII.E
\+{\bR}{\mathbf{R}}                 % "any rotation matrix" in Sec VIII.E
\+{\beps}{\boldsymbol{\epsilon}}    % generator of rotations
\+{\CL}{\hat{\boldsymbol{\mathcal{L}}}} % long linearized operator
\+{\CLd}{\CL^\dagger}               % .., adjoint
\+{\Claytonphase}{\chi}            % "Another way of analysing spiral waves is by computing their phase"
\+{\c}{c}                           % frame speed
\+{\cf}{c_f}                        % front speed
\+{\Domain}{\mathcal{D}}            % spatial domain
\+{\dirac}{\delta}                  % Dirac's delta function
\+{\dist}{d}                        % cycloid step in resonant meander
\+{\E}{E}                           % electric field
\+{\Ecrit}{E_{crit}}                % .., critical value
\+{\ex}{\vec{e}_x}                  % unit vector in x direction
\+{\ffield}{\mathbf{f}}\+{\gfield}{\mathbf{g}} % arbitrary column fields
\+{\ff}{f}                          % any function to be averaged
\+{\findex}{f}                      % phase index in sec VIII.E
\+{\Group}{G}                       % system's symmetry group
\+{\Gone}{\Gamma_1}                 % filament tension, scalar component
\+{\Gtwo}{\Gamma_2}                 % filament tension, pseudoscalar component
\+{\g}{\gamma}                      % element of transformation group
\+{\gg}{g}                          % a function describing transformation between lab and moving frame
\+{\gone}{\gamma_1}                 % filament tension, scalar component
\+{\gtwo}{\gamma_2}                 % filament tension, pseudoscalar component
\+{\HL}{\hat{\mathbf{L}}}           % short linearizer operator
\+{\HLd}{\HL^\dagger}               % .., its adjoint
\+{\HM}{{\mathbf{Q}}}               % matrix of mobilities
\+{\HP}{{\mathbf{P}}}               % diffusion matrix
\+{\h}{h}                           % perturbation component
\+{\hh}{\mathbf{h}}                 % perturbation column
\+{\heta}{\eta}                     % .., magnitude (esp. as small parameter)
\2{\I}{{I^{#1}}_{#2}}               % instant inner product
\+{\J}{J}                           % isotropy subgroup
\+{\Jac}{\mathbf{J}}                % Jacobian of tip -> centre coord transform
\+{\JacH}{\Jac^H}                   % .., Hermitian adjoint
\+{\j}{j}                           % index of a component
\+{\k}{k}                           % index of a component
\+{\l}{\ell}                        % index of a component
\+{\K}{K}                           % sign of $\romgo=\omega$ 
\2{\kron}{{\delta^{#1}}_{#2}}       % Kronecker delta as tensor
\+{\kur}{\kappa}                         % filament curvature
\+{\la}{\lambda}                    % e-val
\+{\lcs}{\epsilon}                  % 2D Levi-Civita; NB \eps^A_{\hs B} vs \eps^{\hs A}_B ?
\2{\lcsf}{{\lcs^{#1}}_{#2}}         % .. "forward" variant
\2{\lcsb}{{\lcs_{#1}}^{#2}}         % .. "backward" variant
\+{\linop}{\hat{\ell}}              % linearized operator in lab space-time
\+{\linopd}{\linop^\dagger}         % .., its adjoint
\+{\MM}{\mathcal{M}}                % the "inertial manifold"
\+{\Att}{\mathcal{A}}                % the attractor in quotient space
\+{\Quot}{\mathcal{Q}}                % the quotient space
\+{\PS}{\mathcal{V}}                % the phase space

\+{\M}{M}\+{\N}{N}                  % collective coordinate indices, comoving frame
\+{\Mb}{\mathbf{M}}                 % permittivity matrix 
\2{\Mel}{{\mathrm{M}^{#1}}_{#2}}    % .. element, raw
\2{\mel}{{\mathrm{m}^{#1}}_{#2}}    % .. element, averaged (?)
\2{\qel}{{\mathrm{Q}^{#1}}_{#2}}    % .. element, averaged (?)
\+{\mint}{\ell}                     % an integer number
\+{\m}{m}\+{\n}{n}                  % collective coord indices, lab frame
\+{\nm}{n}                          % meander period counter 
\+{\mm}{m}                          % .., another one
\+{\mOmgo}{\Omega}                  % .., unperturbed
\+{\mPsi}{\Psi}                     % unperturbed meandering phase
\+{\mpsi}{\psi}\+{\mpsio}{\psi_0}   % meandering phase, its initial value
\+{\Nr}{N}                          % order of resonance in phase-locked drift
\+{\Ns}{N}                          % number of space coordinates
\+{\Nv}{n}                          % number of state variables
\+{\NBS}{N_{BS}}                    % number of broken symmetries/crit modes
\+{\NBSS}{N_{BSS}}                    % number of broken spatial symmetries
\+{\NBTS}{N_{BTS}}                    % number of broken time symmetries (0 or 1)

\+{\romgo}{\omega}                  % .., unperturbed
\+{\somg}{\omega_s}                 % .., of the spiral 
\+{\Phase}{\Phi}                    % tip orientation - not used
\+{\perPhi}{\per\Phi}               % "unperturbed rotation phase"
\+{\perphi}{\per\framephi}          % "correction to rotation phase"
\+{\polang}{\theta}                 % polar angle
\+{\Qty}{A}                         % "an arbitrary quantity" ?
\+{\q}{q}                           % tip velocity in comoving frame (??)
\2{\R}{{R^{#1}}_{#2}}               % rotation tensor
\2{\Q}{{Q^{#1}}_{#2}}               % tension tensor tensor
\+{\RR}{\mathcal{R}}                % right-hand side of the RDS
\+{\r}{\vec{r}}                     % position vector, lab frame
\+{\rt}{r}                          % tip traj radius
\+{\rphi}{\vartheta}\+{\rphio}{\vartheta_0}   % rotation phase of spiral in lab frame; its initial value
\+{\rPhi}{\Phi}\+{\rPhio}{\Phi_0}   % rotation phase of spiral in moving frame; its initial value
\+{\framephi}{\varphi}              % rotation position of the moving frame
\+{\rPhia}{\avg{\rPhi}}     % .., sliding average
\1{\SE}{\mathrm{SE}(#1)}            % special Euclidean group
\+{\SEN}{\mathrm{SE}(\Ns)}            % special Euclidean group
\+{\Subgroup}{H}                    % solutions's spatiotemporal symmetry group
\+{\SSubgroup}{J}                    % solutions's spatial symmetry group
\+{\s}{s}                           % tip speed in Barkley-Hopf bifurcation?
\+{\shift}{\tilde{\varepsilon}}             % any coord shift, esp. as small parameter
\+{\T}{T}                           % any time shift
\+{\To}{T}                          % meandering period, unperturbed
\+{\Trot}{\mathcal{T}}                     % rotation period
\+{\Troto}{\base\Trot}                  % .., unperturbed
\+{\t}{t}                           % time, lab frame
\+{\tuu}{\pert{\uu}}                % the RD column field correction
\+{\UU}{\mathbf{U}}                 % Unperturbed solution in lab FoR
\+{\u}{u}                           % RD components; one component
\+{\uu}{\mathbf{u}}                 % the RD column field
\+{\uuo}{\base\uu}                  % .., unperturbed spiral in the moving frame
\1{\uc}{u^*_{#1}}                   % pinning constant
\+{\V}{V}                           % (generalised) drift velocities
\+{\VV}{\mathbf{V}}                 % right e-func
\+{\vE}{\vec{E}}                    % electric field vector
\+{\vN}{\vec{N}}                    % filament main normal vector
\1{\vo}{\c^{#1}}                   % tip vel in corot frame 
\1{\pv}{v^{#1}}                     % tip velocity perturbation
\+{\WW}{\mathbf{W}}\+{\W}{W}        % left e-funcs, its component
\+{\ww}{\mathbf{w}}\+{\w}{w}        % .., in the lab frame
\+{\X}{X}                           % centre position in index notation; X coord of the same
\+{\Xv}{\vec{\X}}                   % .., vector of
\+{\x}{x}                           % generic position vector in index notation; Cartesian x coord
\+{\xv}{\vec{x}}                    % .., vector notation
\+{\xp}{\rho}                       % comoving/corotating frame coords 
\+{\tp}{\tau}                       % comoving/corotating frame time
\+{\Y}{Y}                           % y-coord of centre position
\+{\y}{y}                           % Cartesian y coord
\+{\z}{z}                           % dummy integration variable

\+{\ctip}{\mathcal{Z}}
\+{\cfil}{\mathcal{X}}
\+{\ptip}{Z} % tip or wavefront spatial coordinate
\+{\pfil}{\X} % filament spatial coordinate
\+{\tfil}{T} % temporal coordinate
\+{\dfil}{\zeta} % displacement tip from filament
\+{\orig}{{O}}
\+{\ccent}{\mathcal{C}}
\+{\pcent}{C}

\+{\NBSS}{\ensuremath{N_{BSS}}}

\+{\SS}{\mathbf{S}}
\+{\eps}{\epsilon} % to be extinguished as ambiguous?

\+{\alpx}{\alpha}                   % some new index ?
\+{\betx}{\beta}                    % some new index to go with $\alpx$
\+{\gamx}{\gamma}                   % .. and another one
\+{\mux}{\mu}                       % .. and another one
\1{\dalp}{m_{#1}}                   % a mysterious new vector introduced in the last subsection. m is overused; change to \mu?
\2{\NN}{{N^{#1}}_{#2}}              % ?
\+{\Mint}{p}                        % a small integer: numerator of a rational 
\+{\Nint}{q}                        % a small integer: denominator of a rational 

\newcommand{\rotleq}{\rotatebox[origin=c]{+90}{\ensuremath{\leq}}}

\usepackage{versions}
\usepackage[usenames,dvipsnames]{xcolor}

\excludeversion{checked}
\markversion{detail}

\begin{document}
%\title{Two regimes of filament tension for meandering scroll waves}
\title{A response function framework for the  dynamics of \\ meandering or large-core spiral waves and modulated traveling waves}

%\author{Hans Dierckx, Alexander Panfilov, Henri Verschelde, Vadim Biktashev and Irina Biktasheva}

\author{H. Dierckx}
\affiliation{Department of Physics and Astronomy, Ghent University, 9000 Ghent, Belgium}
\author{A. V. Panfilov}
\affiliation{Department of Physics and Astronomy, Ghent University, 9000 Ghent, Belgium}
\affiliation{Laboratory of Computational Biology and Medicine, Ural Federal University, Ekaterinburg, Russia}
\author{H. Verschelde}
\affiliation{Department of Physics and Astronomy, Ghent University, 9000 Ghent, Belgium}
\author{V.N. Biktashev}
\affiliation{College of Engineering, Mathematics and Physical Sciences, University of Exeter, Exeter EX4 4QF, UK}
\author{I.V. Biktasheva}
\affiliation{Department of Computer Science, University of Liverpool, Liverpool L69 3BX, UK}
\affiliation{College of Engineering, Mathematics and Physical Sciences, University of Exeter, Exeter EX4 4QF, UK}

\date{\today}

\begin{abstract}
In many oscillatory or excitable systems, dynamical patterns emerge which are stationary or periodic in a moving frame of reference. Examples include traveling waves or spiral waves in chemical systems or cardiac tissue.
We present a unified theoretical framework for the drift of such patterns under small external perturbations, in terms of overlap integrals between the perturbation and the adjoint critical eigenfunctions of the linearised operator (\ie\ `response functions'). For spiral waves, the finite radius of the spiral tip trajectory as well as spiral wave meander are taken into account. Different coordinates systems can be chosen, depending on whether one wants to predict the motion of the spiral wave tip, the time-averaged tip path, or the center of the meander flower. The framework is applied to analyse the drift of a meandering spiral wave in a constant external field in different regimes. 
\end{abstract}

\maketitle

%Winfree:1972
% \tableofcontents
% \newpage
\section{Introduction \label{sec:introduction}}

Spiral waves are remarkable self-sustained patterns which arise in various extended systems, including oscillating chemical reactions \cite{Jahnke:1988}, catalytic oxidation \cite{Nettesheim:1993} and biological systems. In biology, spiral waves have been observed  across vastly different spatial scales, as they organise intracellular calcium waves \cite{Lechleiter:1991}, slime mould aggregation \cite{Siegert:1992}, waves of spreading depression in the brain \cite{Gorelova:1983}, and cardiac contraction during arrhythmic events \cite{Allessie:1973,
 Gray:1998, Witkowsky:1998,  Haissaguerre:2014}. In the context of cardiac arrhythmias, different drift regimes are thought to correspond to different heart rhythm disorders \cite{Gray:1998}. Moreover, tracking of rotation centers of cardiac rotors was shown to improve ablation therapies of atrial arrhythmias~\cite{Narayan:2012}.
% The prediction of spiral wave drift and stationary solutions in anisotropic tissue has so far been focussed on the circular-core case \cite{Rogers:1994, Wellner:2000, Wellner:2002, Davydov:2004, Verschelde:2007}.

The development of asymptotic description of drift of spiral waves has been contingent on two important issues:  
localization of adjoint symmetry modes~\cite{ Biktasheva:1998,
  Biktasheva-Biktashev-2001,   Hakim:2002, Biktasheva:2009,
  Marcotte:2015, Dierckx:2017PRL}
and an appropirate choice of the drift system of reference~\cite{
  Keener:1988,
  Biktashev:1994,
  Biktasheva:2003,
  Biktasheva-etal-2010}. On this pathway, the choice of an ``optimal'' system of reference naturally depends on the unpertubed spiral wave dynamics, \ie\ whether it is (i) a rigidly rotating spiral wave with its tip following a perfect circle, (ii) a bi-periodic ``meander'' regime when the spiral wave periodically changes its shape and its tip describes a bi-periodic flower-like trajectory, or (iii) more complicated cases dubbed `hypermeander'.
While the asymptotic theory of the drift of rigidly rotating spiral waves is the most advanced~\cite{
  Biktasheva:2003, Biktasheva-etal-2010}, the theory of hypermeander drift is the one least developed~\cite{ Biktashev:1998,Nicol-etal-2001,  Ashwin-etal-2001}. The current work aims to lay out a technical framework for the drift of a bi-periodic meander regime, and illustrate its application on a simple example of electroforetic drift.

\begin{figure*}
\raisebox{3cm}{a)} \includegraphics[width=0.28 \textwidth]{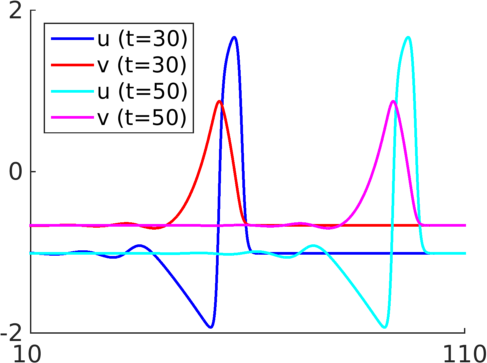}
\raisebox{3cm}{b)}
\begin{tabular}[b]{c}
\includegraphics[width=0.3 \textwidth]{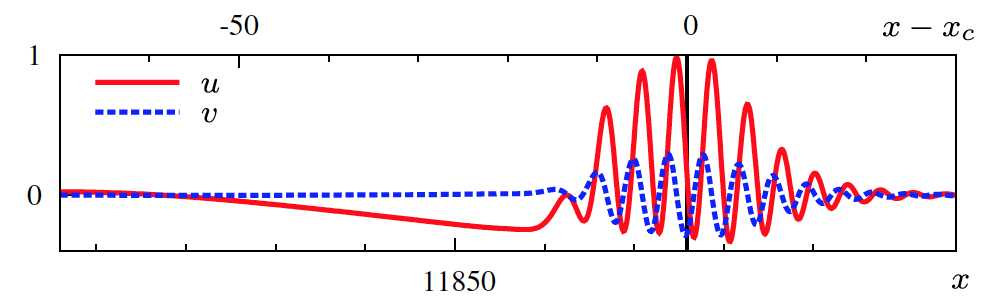}\\
\includegraphics[width=0.3 \textwidth]{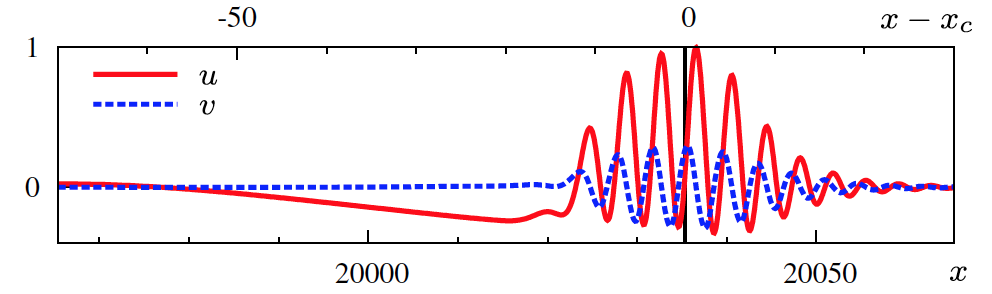}
\end{tabular}
\raisebox{3cm}{c)} \includegraphics[width=0.3 \textwidth]{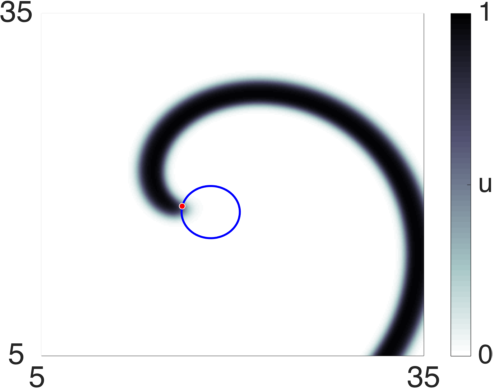}\\
\raisebox{3cm}{d)} \includegraphics[width=0.3\textwidth]{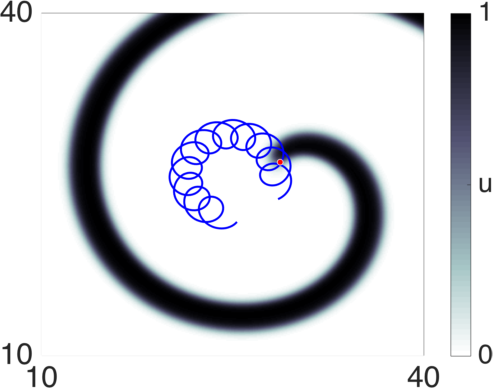}
\raisebox{3cm}{e)} \includegraphics[width=0.3\textwidth]{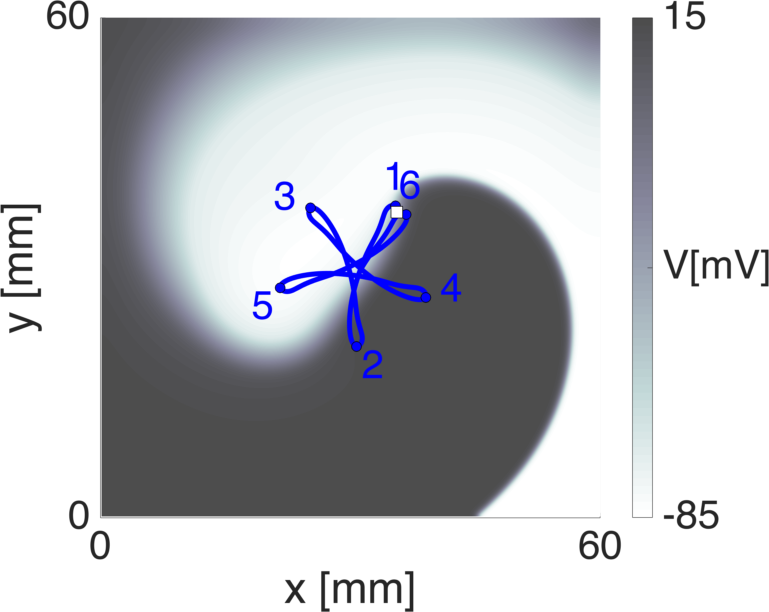} 
\raisebox{3cm}{f)} \includegraphics[width=0.3\textwidth]{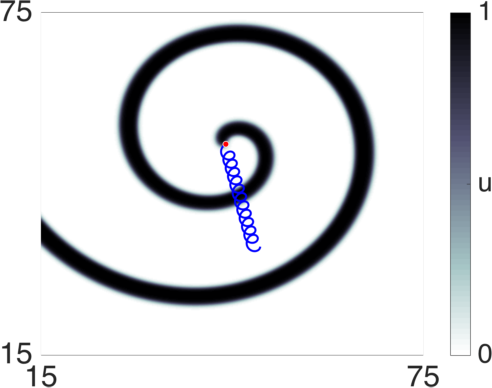}
\caption[]{Qualitatively different one- and two-dimensional patterns that arise as solutions to the reaction-diffusion equation \eq{RDE}, depending on the chosel reaction model $\bF(\uu)$ and initial conditions. a) Stationary traveling wave in the FitzHugh-Nagumo model \cite{FitzHugh:1961} ($\fhneps = 0.3$, $\fhnbet=0.68$, $\fhngam=0.5$). b) Modulated 1D traveling wave in the FitzHugh-Nagumo model with cross-diffusion, shown at $\t=3000$ and $\t=5000$; reproduced from \cite{Biktashev:2011}. c) Rigidly rotating spiral wave in Barkley's model \cite{Barkley:1991} ($\apar=0.52$, $\bpar=0.05$, $\cpar=0.02$). d) Spiral wave with Barkley kinetics exhibiting flower-like meander with inward petals ($\apar=0.58$, $\bpar=0.05$, $\cpar=0.02$). e) Meandering spiral wave with a linear-core in the Fenton-Karma Guinea Pig cardiac tissue model \cite{Fenton:1998}. Labels indicate the order in which petals are visited. 
f) Spontaneous spiral wave drift in the case of resonant meander, in Barkley's model ($\apar=0.625$, $\bpar=0.05$, $\cpar=0.02$). \label{fig:patterns} }
\end{figure*}

The propagation of cardiac excitation, as well as many other systems in which spiral waves are observed, can be described by reaction-diffusion (RD) systems (see, \eg~\cite{Clayton:2011}), 
\alnn{
  \dd_t \uu = \dd_{\j} \left[ \HP^{\j\k}(\r,\t) \dd_{\k} \uu \right] + \bF(\uu,\r,\t) . 
}
Here, $\uu=\uu(\r,\t)\in\Real^\Nv$ is the list (column-vector) of $\Nv$ state variables, 
$\uu=\Mx{\u_1,\dots,\u_\Nv}^T$, 
\eg\ the concentrations of reacting species, 
$\r=(\x^\j)\in\Domain\subset\Real^\Ns$ is the vector of Cartesian coordinates, $\Ns=1,2$ or 3, 
and $\t$ is time. 
The vector-function $\bF:\Real^{\Nv}\times\Real^{\Ns+1} \rightarrow  \Real^{\Nv}$ describes the local dynamics of state variables, \eg\ the reaction rates.
The space-time dependent matrix-tensor $\Mx{\HP^{\j\k}} : \Real^{\Ns+1} \to \Real^{\Nv\times\Nv}\times\Real^{\Ns\times\Ns}$ contains the diffusivities of the state variables on its diagonal (of which some may be vanishing). 
Non-diagonal terms of $\HP$ are present in systems with cross-diffusion; see \eg\ Ref.  \cite{Biktashev:2011} and Fig. \ref{fig:patterns}(b). 

The asymptotic description follows the paradigm that first an idealized mathematical situation is considered, which is an unbounded, perfectly stationary, homogeneous and isotropic medium governed by the isotropic reaction-diffusion equation (RDE)
\al{
 \dd_t \uu = \HP \Lap \uu + \bF(\uu) \eqlabel{RDE}
}
and then any deviations from the idealized picture are considered perturbatively.
Here the diffusion matrix $\HP$ is constant and the kinetic functions
$\bF(\uu)$ are taken uniform in space and time. In the context of pattern formation, a bistable, oscillatory or
excitable point system is commonly used.  For well-chosen sets of
reaction kinetics, \ie\ functions $\bF(\uu)$, the system \eq{RDE} may
sustain waves of finite amplitude, which propagate through the medium,
see Fig. \ref{fig:patterns}(a--b).

Although our methodology works for any reaction-diffusion system allowing pattern formation, we will use below examples with the reaction kinetics of FitzHugh-Nagumo \cite{FitzHugh:1961, Biktasheva:2009} 
\alnn{
\HP &= \diag(1,0), \\
\bF(\uu) &= \left[\fhneps^{-1} (\u_1-\u_1^3/3 - \u_2), \quad \fhneps (\u_1 -\fhngam \u_2 + \fhnbet) \right]^T,
} 
Barkley \cite{Barkley:1991} 
\alnn{
\HP &= \diag(1,0), \\
\bF(\uu) &= \left[\cpar^{-1} \u_1(1-\u_1)[\u_1-(\u_2+\bpar)/\apar], \quad \u_1-\u_2 \right]^T
}
and Fenton and Karma. The equations and parameters of the latter model (with 3 state variables) are found in \cite{Fenton:1998}. 

% 1D waves, %2D spirals with circular core
In $\Ns\geq 2$ spatial dimensions, the reaction-diffusion equation \eq{RDE} may sustain rotating spiral-shaped solutions, as illustrated in Fig. \ref{fig:patterns}(c--f). Spiral waves are commonly characterised by their tip trajectory, which can be found in different ways \cite{Zykov:1997, Fenton:1998}, \eg\ as the locus where 
\al{
  \u_\j(\x,\y,\t) = \uc\j, \qquad  \u_\k(\x,\y,\t) = \uc\k. \eqlabel{tipline}
}
where $\j$, $\k$ are indices of two selected state variables, $\j\ne\k$, and $\uc{\j}, \uc{\k}$ are two appropriately chosen constants. 
In $\Ns=2$ spatial dimensions, Eq. \eq{tipline} means the intersection of isolines of two selected field components, where one isoline may correspond to a threshold level of the activator field, and the second condition separates the ``front'' and ``back'' parts of the first isoline based on the values of the inhibitor field. E.g. for Barkley kinetics, we use $\j=1, \uc{\j}=0.5, \k=2, \uc{2} = 0.5a-b$. It is often also convenient to take $\dd_\t\u_\j=0$ as the second condition which corresponds to a point where the velocity of the activator isoline vanishes.
Example tip trajectories are shown in Fig. \ref{fig:patterns}(c--f). Another way of analysing spiral waves is by computing their phase, \eg\ defined as a polar angle $\Claytonphase$ in a suitably chosen plane in the phase space of local kinetics:
\al{
 \tan \Claytonphase(\r,\t) = \frac{\u_\k(\r,\t) - {\uc\k}}{\u_\j(\r,\t) - {\uc\j}} , \eqlabel{ps}
}
This mapping will produce a phase singularity (PS) in the vicinity of the spiral tip \cite{Clayton:2005}. If $\j,\k, \uc\j, \uc\k$ are set to the same values in Eqs. \eq{tipline} and \eq{ps}, the PS coincides with the spiral wave tip definition. Otherwise, different observers will generally not exactly agree on the tip or phase singularity position \cite{Gray:2009}.
 
% 3D
In three spatial dimensions, the  solution to the reaction-diffusion equation \eq{RDE} consisting of a stack of spiral waves is known as a `scroll wave'. When tracking the scroll wave's tip or PS, Eqs. \eq{tipline} and \eq{ps} produce a curve that is known as the `scroll wave filament'. The same remark as in 2D holds here: different algorithms  and chosen thresholds will yield different filament curves, which generally lie in each other's vicinity, \ie\ in the tubular scroll wave core region. 

% circular core
However, if the tip trajectory is circular, all tip paths and PS trajectories will in 2D describe circles around a unique rotation center $\ccent$ at the same angular frequency $\romgo$. In previous works, the dynamics of circular-core spiral waves have been analysed in terms of the motion of the rotation center $\ccent$ \cite{Verschelde:2007, Biktashev:2010, Dierckx:2012, Dierckx:2013}. 
For circular tip paths, the region inside it never excites and is sometimes called the `spiral wave core'. The dynamics of circular-core spiral waves is reasonably well understood, and their drift response to small external stimuli can be found by projecting the stimulus on the inertial manifold of the system using so-called `response functions' \cite{Keener:1988, 
Biktashev:1995b, Biktasheva:2003, PanfilovDierckx:2017}.

% meander
Interestingly, in several experiments and numerical simulations, both in chemical \cite{Winfree:1973,
  Steinbock:1993} and biological \cite{Haissaguerre:2014, Yamazaki:2012, Efimov:1995, Courtemanche:1998, Biktashev:1996model, Fenton:1998, BuenoOrovio:2008} systems, the spiral wave was  found to perform not a rigid rotation, but a more complicated motion, a phenomenon known as `spiral wave meander'~\cite{Roessler-Kahlert-1979,Zykov-1986,Winfree:1991}.  In excitable models, meander can arise due to the wave front interacting with the wave back from the previous excitation. In the simpler cases, the meander is quasi-periodic, in the sense that the evolution is periodic up to a Euclidean transformation. This transformation can always be written as a pure rotation or a pure translation (see below), which is why one can refer to it as `biperiodic' meander, even when the tip trajectory is not a superposition of two circular motions. In more complicated cases, often called ``hypermeander''~\cite{Winfree:1991}, the tip motion can have more than two periods~\cite{Nicol-etal-2001} or be chaotic, resulting in deterministic Brownian motion of the tip position \cite{Biktashev:1998}; we will not consider such cases here. 

A significant step in understanding meander of spiral waves was made by Barkley et al. \cite{Barkley:1990b} and Karma~\cite{Karma-1990}, who showed that the transition to meander from the circular-core case exhibits typical features of a Hopf bifurcation. 
Barkley~\cite{Barkley:1994} then showed that the process can be described by a set of five coupled ordinary differential equations. As a result, a tip trajectory is formed that is a superposition of two circular rotations, making a flower-like tip trajectory. This picture was put on constructive footing in \cite{Wulff-1996,Fiedler-etal-1996,Golubitsky:1997,Sandstede:1997},
using ``skew-product'' representation of the reaction-diffusion system exploiting its symmetry with respect to Euclidean motions of the plane. These works described the equivariant Hopf bifurcation (\ie\ the regime close to the transition to meander), while many chemical and cardiac models show a qualitatively different tip trajectory, \ie\ zig-zagged, star-like path known also as the ``linear core'' case. It turned out, however, that skew-product representation does not need to be restricted to the vicinity of the transition zone, and the more complicated meander pattern can be described as relative periodic solutions not necessarily related to a Hopf bifurcation~\cite{Biktashev:1996,Foulkes:2010}. 

Correspondingly, evolution of biperiodically meandering spirals, both flower-like and star-like, in response to small perturbations, can be analysed by linearising the system around a relative periodic orbit \cite{Dierckx:2017PRL}. To show the steps of this process in detail is the purpose of this manuscript; it extends the procedure for rigidly rotating spirals found in \eg\ \cite{Biktashev:1995b,Biktasheva:2003,Biktasheva:2009}. Since periodically deforming waves in one spatial dimension are obtained as a special case, we also describe their dynamics. 

% binding text? 
% laws of motion and time-averaging

A significant part of this paper will be dealing with introducing different coordinate systems that are suitable to capture the drift dynamics of solutions to the RDE \eq{RDE} because, depending on which definition of filament one adopts one can obtain simple or complex laws of motion. These laws can be further simplified if one averages in time over rotation cycles of the spiral or scroll waves~\cite{Biktashev:1995b}. For non-stationary spiral or scroll waves, a possible averaging method is to analyse the motion in a Fourier series and only keep the non-oscillating, secularly growing terms \cite{Dierckx:2013virtual, Dierckx:2017}. Some recent works have computed the leading order eigenmodes in the laboratory frame of reference \cite{Marcotte:2015, Marcotte:2016}. At the end, different descriptions need to be compatible, and it is our aim to illustrate that they all describe the same dynamics. 

This paper is organised as follows. In Sec. \ref{sec:symmetries} we discuss the concept of symmetry breaking by the formed patterns, which will determine the number of degrees of freedom required to uniquely describe such patterns. We first illustrate this concept in Sec. \ref{sec:1D} on traveling waves in one dimension, treating the cases of rigidly moving and periodically deforming waves. In Sec. \ref{sec:2D} we consider spiral wave patterns in two spatial dimensions, in the regimes of rigid rotation, meander and resonant meander. We consider in detail multiple possible frames of reference, which are defined with respect to either instantaneous or time-averaged dynamics. In Sec. \ref{sec:linear} we linearise around the solution and discuss the critical eigenmodes of the system. This knowledge is used in Sec. \ref{sec:drift} to derive the equations of motion for modulated wave patterns, in a manner that is valid for both traveling and spiral waves.  In Sec. \ref{sec:geom} we re-interpret
  our approach in terms of the geometry of the phase space of the reaction-diffusion system considerd as a dynamical system. This part is optional, but we found it useful to include as few works make the connection between the `physics' style of description and the more abstract dynamical systems viewpoint.

In Sec. \ref{sec:applications} the drift response of a spiral wave in a constant external vector field is analysed. We show how the dynamics can be averaged over rotation and temporal phases in order to find a simpler, manageable equation of motion. In particular, we demonstrate that if a meandering spiral wave is phase-locked to a constant external field of magnitude $E$, its drift velocity can be larger than order $E$: it is close to the `orbital velocity' which which the spiral wave circumscribes the meander flower. 

We conclude with a discussion of our present work in Sec. \ref{sec:discussion}. To assist the reader in remembering the notations used, we provide a list of notations in the Appendix.

\section{Symmetry considerations} \label{sec:symmetries}

It is customary to consider Eq. \eq{RDE} in a bounded spatial domain with Neumann boundary conditions, which agrees with many physical situation where such models are relevant, and this is used in all numerical simulations of this work. However, to develop the theory of symmetry-breaking, we here consider the whole space $\Real^\Ns$, with {$\Ns=1$} for wave fronts and {$\Ns=2$} for spiral waves. 
The two sets of boundary conditions do not need to conflict, as numerical evidence in various reaction-diffusion patterns \cite{ Biktasheva:1998,
  Biktasheva-Biktashev-2001,   Hakim:2002, Biktasheva:2009,
  Marcotte:2015, Dierckx:2017}
 shows that the sensitivity of patterns to external stimuli, including boundary conditions, is strongly localised. The mathematical condition for this is that the adjoint critical eigenfunctions (`response functions' \cite{Biktasheva:2003}) $\WW^\M$ defined in Eqs. \eqref{defW} below
decay exponentially far from the wave front (in 1 spatial dimension) or far from the spiral wave tip (in 2 spatial dimensions). Therefore, faraway boundaries are not felt at \eg\ the wave front or spiral tip.

Since the system~\eq{RDE} involves the spatial variables only via the Laplacian $\Lap$ in the whole space $\Real^\Ns$, it is equivariant (symmetric) with respect to the isometric transformations $\g:\Real^\Ns\to \Real^\Ns$ of this space, meaning that if $\u(\r,\t)$ is a solution, then ${\trans\u}(\r,\t)=\u(\g^{-1}\r,\t)$ is also a solution, for any isometry $\g$. For our purposes it is sufficient to consider only the orientation-preserving transformations, that is, $\g\in\SE\Ns$. Besides, being autonomous, this system is equivariant with respect to the group of translations of the time axis, $\SE1\sim\Real$, that is if $\u(\r,\t)$ is a solution, then ${\trans\u}(\r,\t)=\u(\r,\t-\T)$ is also a solution for any $\T\in\Real$.  (Note that this is different from saying the solution is `invariant' with respect to time translation, which would imply a solution that is constant in time.)
Thus, we can say overall that \eq{RDE} is symmetric with respect to the inhomogeneous Galilean group $\Group=\SE\Ns\times\Real$. The dimensionality of this Lie group is $\dim(\Group)=\Ns(\Ns+1)/2+1$. 

%When an element of $\Group$ acts on a state $\uu(\r,\t)$, there are three possibilities: one obtains either the same state, a state of the same solution (orbit) at a different time, or a different solution. In the first case, the state is invariant under the symmetry. This happens \eg\ with a plane wave in $\Ns =2$, which is invariant for translations along the wave front. This case can be reduced to the study of a traveling wave in $\Ns=1$ spatial dimension. 

Any given solution $\u(\r,\t)$ of Eq. \eq{RDE} may be invariant with respect to none, or all, or part of the symmetries of the equation itself. Invariance of the solution with respect to a purely spatial symmetry $\g \in \SEN$ can in general be reduced to a lower-dimensional solution and is not further considered. E.g. a plane wave in $\Ns =2$, which is invariant for translations along the wave front. This case can be reduced to the study of a traveling wave in $\Ns=1$ spatial dimension. An interesting case arises when a solution is invariant with respect to a spatiotemporal symmetry. E.g. for a rigidly translating traveling wave in $\Ns=1$ at constant speed $\c$, $\uu(\r + \c \T ,\t + \T) = \uu(\r, \t)$ for any $\T$, which implies that there are members of $\Group$ that leave the solution invariant. 

Let the group of symmetries leaving a solution invariant be $\Subgroup$, $\Subgroup\le\Group$. If this is also a Lie group, then application of all possible transformations $\g\in\Group$ to  $\u(\r,\t)$ will produce a manifold of solutions, of dimensionality $\NBS=\dim\Group-\dim\Subgroup$. We refer to this dimensionality as the number of broken symmetries (implying that the ``unbroken ones'' are in the subgroup $\Subgroup$). By construction, the tangent vectors to this manifold of solutions are solutions of the \eq{RDE} linearised on $\u(\r,\t)$ which do not exponentially grow or decay with time, so they make the center subspace of this solution. If the solution is an attractor to the system, this manifold of solutions will be called an `inertial manifold', see details in Sec. \ref{sec:geom}. 
%
% "center subspace" is tangent to the manifold, so it is not the same! 
%

% called `intertial manifold` below. 

In the following text, it will sometimes be convenient to indicate for every quantity, \eg\ wave front or tip coordinates, the reference frame in which it is defined, using different math accents.  We shall call the `simplest' frame that makes the solution constant or periodic the \emph{center frame of reference}.  For any quantity $\ff$, its value pertaining to this frame of reference, will be denoted as $\cen{\ff}$.  The frame which best follows the wave front or spiral wave tip will be called the 'tip frame`, with the corresponding notation $\tip{\ff}$. An list of all notations and symbols used throughout this manuscript is given in the appendix. 

\section{One-dimensional traveling waves \label{sec:1D}}

\subsection{General framework}
Let us first consider a generic choice for the frame of reference to describe a traveling wave $\UU(\x,\t)$.  We denote the origin of the lab frame with Cartesian coordinate $x$ as $\orig$. In the lab frame, we can define the wave front position $\x=\ctip(\t)$ by thresholding a state variable $\u_\j(\ctip(\t),\t) ={\uc\j}$ and demanding that $\dd_\t\u_\j(\ctip(\t),\t)>0$ to distinguish it from the wave back. This is one possible definition for the wave front position as a moving point $\ctip(\t)$, with lab frame coordinate $\x = \ptip(\t)$. However, alternative definitions are possible, \eg\ for the wave shown in Fig. \ref{fig:patterns}(b) the barycentre of $|\u_2|^2$ was used to denote wave position. 

\begin{table*}
\begin{tabular}{|c|c|c|c|c|c|c|c|} \hline
traveling wave type & 
$\NBSS$ & 
$\NBTS$ & $\NBS$ &
Frame name &
Origin & 
Choice & 
Frame velocity \\\hline\hline
%----------
any & 
1 & 0 or 1 & 1 or 2 &
lab  &
fixed & 
$\cfil = \orig$ & 
$\c=0$ \\  
\hline\
rigidly moving front &
1 &  0 & 1&
co-moving & 
at front & 
$\cfil = \ctip$ & 
$\c = \cf$  \\ \hline 
%----------
\multirow{2}{*}{periodically modulated front} & 
\multirow{2}{*}{1} & \multirow{2}{*}{1} & \multirow{2}{*}{2}  &
fully co-moving &
at front & 
$\cfil = \ctip$ & 
$\c(\mPsi) = \cf(\mPsi)$  \\
%----------
& 
 & &&
constant-speed co-moving frame &
near front & 
$\cfil \ne \ctip$ & 
$\c = \avg{\cf(\mPsi)}=\const$  \\ \hline 
\end{tabular}
\caption{Overview of reference frames used in the text to describe the motion of 1D traveling wave patterns. $\NBSS$: number of broken spatial continuous symmetries, $\NBTS$: number of broken temporal continuous symmetries,
$\NBS = \NBSS + \NBTS$: number of broken continuous symmetries, 
 $\orig$: lab frame origin, $\ctip$: instantaneous wavefront position, $\cfil$: chosen origin of the moving frame, $\cf$: speed of the wave front in the lab frame, $\mPsi$: temporal phase describing the progression of the solution along a deformation cycle (relative periodic orbit). \label{tab:1D}}
\end{table*}

Next, we introduce a (yet unspecified) moving frame of reference with origin $\cfil$ that has lab frame coordinate $\pfil(t)$ and denote the new spatial coordinate in the moving frame as $\xp$:

\al{
\xp &=  \x - \pfil(\t) ,&  \qquad \t &= \tp.  \eqlabel{comoving}
}

The moving frame can be chosen in different manners; an overview is given in Tab. \ref{tab:1D}. 

The instantaneous wave front position is denoted $\ptip(\t)$, it is found at the position $\xp = \dfil(\t)$ in the moving frame:
\alnn{
 \ptip(\t) = \pfil(\t) + \dfil(\t).
}

\subsection{Symmetry breaking}

In a one-dimensional homogenous spatial domain ($\Ns=1$), the RDE \eq{RDE} is invariant with respect to translation in time and space, \ie\ $\Group\sim\Real\times\Real$, and $\dim(\Group)=2$.
A traveling-wave solution does not have translational symmetry, neither in time nor space, so we can say it breaks the translational symmetry. 
However, if the wave travels with constant speed $\cf$ without changing its shape, then 
we can say that $\u(\x-\X,\t-\T)=\u(\x,\t)$ as long as $\X$ and $\T$ are related by $\X=\cf\T$. That means, that the solution is invariant with respect to a one-parametric group of transformations, so $\dim\Subgroup=1$, and $\NBS=1$. 
% The next question is: is the time-symmetry also broken? 
% To answer this question, we consider the system disregarding the broken spatial symmetry; one could say we either put ourselves in a co-moving frame of reference or in the quotient system of the dynamics (see Sec. \ref{sec:geom}). If, in the co-moving frame, the solution is invariant in time, there is no additional broken symmetry, bringing the number of broken symmetries $\NBS$ to $1$.  
This is the case of rigidly translating waves, to be treated in subsection
 \ref{sec:rigidwave}. 
However, if there is no frame of reference moving with constant speed in which the solution is constant in time, then the subgroup $\Subgroup$ is trivial or at most discrete, so $\dim\Subgroup=0$, and $\NBS=2$. The case where this dynamics is periodic, so $\Subgroup$ consists of translations $\x\to\x+\mint\cf\To$, $\t\to\mint\To$, $\mint\in\Zahlen$, is treated in subsection \ref{sec:MTW}.
The non-periodic case, when $\Subgroup$ is trivial, falls outside our present scope. 

\subsection{Rigidly translating waves \label{sec:rigidwave}}

Firstly, if the wave is rigidly moving, $\ptip(\t) = \ptip(0) + \cf\t$ with $\cf$ the constant traveling wave speed. In this case, we simply take $\cfil \equiv \ctip$ and $\pfil \equiv \ptip$ and speak of the `co-moving frame'. In this frame, the traveling wave is a solution to the RDE \eq{RDE} of the form: 
\algrp{u0wave}{
  \UU(\x,\t) &= \uuo(\x - \pfil(\t)),    \eqlabel{u0wave-UU}\\
  \dd_\tp \pfil &= \c.            \eqlabel{u0wave-dtX}
}

Note that Eq. \eq{u0wave-UU} describes the type of solution %, \eq{u0wave-rho} is a change of coordinates $\x \rightarrow \xp$ in which the dynamics becomes simpler, 
and \eq{u0wave-dtX} gives the equation of motion for the collective coordinate $\pfil(\t)$ (wave front position) that is introduced since the pattern breaks the translational symmetry along $\x$. If a perturbation of order $\heta$ is added to the system, Eqs. \eq{u0wave-UU} and \eq{u0wave-dtX} will gain additional terms of $\OO(\heta)$ in their right-hand sides, describing the wave profile change and wave speed correction, respectively. 

\subsection{Periodically modulated waves \label{sec:MTW}}

Traveling wave may also possess a variable shape and variable propagation speed $\cf$, in which case $\pfil(\t) = \pfil(0) + \int_0^\t \cf(\t')\,\d\t'$. This situation has been encountered both in experiment and simulations \cite{Manz:2006, Biktashev:2011}, see Fig. \ref{fig:patterns}b for an example. We only treat periodically modulated waves here, where $\cf(\t)$ has temporal period $\To$. In the presence of perturbations, it will be convenient to rescale the period to $2\pi$, define the positive constant $\mOmgo = 2\pi/\To$ and call $\mPsi = \mOmgo\t$ the temporal phase of the solution, which labels how far the wave front shape has gone through its cycle. That is, $\mPsi$ is essentially a scaled time variable, and $\mOmgo$ is the scaling constant.

Throughout this work, we will use the bar notation to denote the average over one temporal cycle: 
\alnn{
  \avg{\ff} = \frac{1}{2\pi}\int_0^{2\pi} &\ff(\mPsi) \,\d\mPsi % \eqlabel{avg}
}
for any $2\pi$-periodic $\ff(\mPsi)$. We now have multiple options to choose the frame of reference. 

The traveling wave solution is now periodic in a moving frame, \ie\ 
\aleq{u0deformingwave}{
 \UU(\x,\t) &=  \uuo(\x - \pfil(\t), \mPsi(\t)), \\
 \dd_\t \pfil &= \c(\mPsi), \\
 \dd_\t \mPsi &= \mOmgo.
}
 
Due to the scaling of time, $\uuo$ is $2\pi$-periodic in $\mPsi$. 

\subsubsection{Fully co-moving frame}

In the `front frame', we take again $\cfil \equiv \ctip$ and $\pfil \equiv \ptip$, at the expense of a non-constant propagation speed $\cf(\mPsi)$.

\subsubsection{Constant-speed co-moving frame}

Alternatively, one may choose to let the frame move at the time-averaged velocity 
\alnn{
  \c = \avg{\cf(\mPsi)} .
}
 
Since $\c$ is constant here, this frame is simpler than the fully co-moving frame. Keep in mind, however, that $\pfil$ now describes the average progression of the wave front position; the true wave front position $\ptip$ can be found from
\alnn{
  \ptip = \pfil +\dfil, \qquad \dfil(\mPsi) =  \int_0^\mPsi \left(\cf(\mPsi') - \c\right) \,\d\mPsi'.  
}

\section{Spiral waves \label{sec:2D}}

\begin{table*}
\begin{tabular}{|c|c|c|c|c|c|c|c|} \hline
Spiral type & 
$\NBSS$ & 
$\NBTS$ &
Frame name &
Origin of moving frame & 
Choice & 
Rotation &
Translation \\\hline\hline
%----------
any & 
3 & 0 or 1& 
lab & 
fixed & 
$\cfil = \orig$ & 
$\romgo = 0$ &
$\vo\a = 0$ \\\hline
%----------
\multirow{2}{*}{\begin{tabular}{c} rigidly \\  rotating \\ \end{tabular}} &
\multirow{2}{*}{3} & 
\multirow{2}{*}{0} & 
center frame &
middle of tip traj. & 
$\cfil = \ccent$ & 
$\romgo = \K\somg$ &
$\vo\a = 0$ \\
%----------
&
&
&tip frame &
at tip & 
$\cfil = \ctip$ & 
$\romgo = \K\somg$ & 
$\vo\A=\const\neq 0$ \\\hline
%----------
\multirow{4}{*}{\begin{tabular}{c}non- \\ resonant \\ meander\end{tabular}} &
\multirow{4}{*}{3} &
\multirow{4}{*}{1} &
center frame &
middle of flower & 
$\cfil = \ccent $ &
$\romgo = \alpo/\To$ &
$\vo\a = 0$ \\
%----------
&
& 
&minimally rotating finite-core frame &
on a circle &
$\cfil \neq \ccent, \ctip$ &
$\romgo = \alpo /\To$& 
$\vo\A=\const\neq 0$ \\
%----------
&
& 
&co-rotating finite-core frame &
on a circle &
$\cfil \neq \ccent, \ctip$ &
$\romgo = \alps/\To $ & 
$\vo\A=\const\neq 0$ \\
%----------
&
&
&tip frame & 
at tip & 
$\cfil = \ctip$ &
$\avg{\romgo(\mPsi)} = \alps/\To$ &
$\vo\A(\mPsi)$  \\\hline
%----------
\multirow{3}{*}{\begin{tabular}{c} resonant \\ meander\end{tabular}} &
\multirow{3}{*}{3} &
\multirow{3}{*}{1} &
minimally rotating co-moving frame &
on a line &
$\cfil \neq \ctip $ &
$\romgo=0$ &
$\vo\A=\const\neq 0$ \\
%----------
&
&
&co-rotating, co-moving frame &
on a line &
$\cfil \neq \ctip $ &
$\romgo=\K\mOmgo$ &
$\vo\a=\const\neq 0$ \\
%----------
&
&
&tip frame &
at tip &
$\cfil = \ctip$ &
%$\romgo(\mPsi)$, $\avg{\romgo} = \K\mOmgo$ &
$\avg{\romgo(\mPsi)} = \K\mOmgo$ &
$\avg{\vo\A(\mPsi)}=0$ \\ \hline 
\end{tabular}
\caption[]{%
Overview of reference frames used in the text to describe the motion of rotating spiral wave patterns. %
$\NBSS$: number of broken continuous spatial symmetries, %
$\NBTS$: number of broken continuous temporal symmetries, %
$\orig$: lab frame origin, %
$\ccent$: middle of the tip trajectory, %
$\ctip$: spiral wave tip, %
$\cfil$: chosen origin of the moving frame (``filament"),  %
$\vo\a$: velocity of $\cfil$ in the lab frame, %
$\vo\A$: its velocity in the rotating frame, $\K= \pm 1$ is the sign of the (period-averaged) frame rotation rate $\romgo$, introduced so that $\somg>0$ or $\mOmgo >0$, respectively. $\NBSS + \NBTS = \NBS$, the total number of broken symmetries.  %
}
\label{tab:2D}
\end{table*}

\subsection{General framework}

Let $\x^\a$, $\a\in\{1,2\}$ be Cartesian coordinates in the lab frame with origin $\orig$. In the lab frame, the spiral wave tip $\ctip$ as defined by \eq{tipline} describes a path $\x^\a = \ptip^\a(\t)$ that may be circular or flower-like, see Fig.~\ref{fig:patterns}. 

To capture spiral wave dynamics, we will introduce a moving frame with origin $\cfil(\t)$, whose lab frame coordinates are $\x^\a = \pfil^\a(\t)$. At all times, the orientation of the new frame of reference with respect to the lab frame is given by the rotation phase $\rPhi(\t)$, with rotation rate $\dd_t \rPhi = \romgo$ that may be time-dependent. So, $\romgo$ takes positive values for counterclockwise frame rotation and negative values for clockwise frame rotation. If $\romgo$ is constant, we introduce $\K = \sgn(\romgo)$; if it is periodic, $\K = \sgn(\avg{\romgo})$. 
Remark that for meandering spirals, $\romgo$ coincides with the spiral rotation rate only if the `tip frame' (defined below) is chosen. For the other frame choices in Tab. \ref{tab:2D}, $\romgo$ equals the (typically slow) precession rate of the meander pattern.   

The spatial coordinates in the moving frame are denoted $\xp^\A$, and co-moving time coordinate is $\tp$. The generic coordinate transformation between the frames is thus given by:
\alnn{
  \x^\a = \pfil^\a + \R\a\A (\rPhi) \xp^\A, \qquad \t = \tp. 
} 
The rotation matrices connecting lab frame coordinates $\x^\a$ to rotating coordinates $\xp^\A$ are given by
\alnn{
 \Mx{ \R\A\a(\rPhi) } &= \Mx{
  \cos \rPhi \ & \sin \rPhi \\
  -\sin \rPhi & \cos \rPhi
  }, \\
  \Mx{ \R\a\A(\rPhi) } &= \Mx{
    \cos \rPhi & -\sin \rPhi \\
    \sin \rPhi & \cos \rPhi
  }.
}
 
Since the generator of rotations is the Levi-Civita symbol $\lcsf\A\B$ ($\lcsf12 = - \lcsf21 = 1$ and zero otherwise), 
it commutes with rotation matrices, and
\alnn{% eq{dphiR}{}
 \dd_\rPhi \R\A\a(\rPhi) &= - \R\A\a(\rPhi) \lcsf\a\b = - \lcsf\A\B \R\B\a(\rPhi), \\
 \dd_\rPhi \R\a\A(\rPhi) &= - \R\a\A(\rPhi) \lcsf\A\B = - \lcsf\a\b \R\b\A(\rPhi). 
}

We will write the polar angle in the chosen frame of reference as $\polang$, while the spatial orientation of the reference frame will be the rotation phase $\rPhi$.
Due to rotational invariance of the RDE \eq{RDE}, the solution will generally depend only on 
\alnn{
  \rphi = \polang - \rPhi.
}
This relation is the angular equivalent of Eq. \eq{comoving}.  

In the moving frame, the spiral wave tip position is found at position $\xp^\A = \dfil^\A(\tp)$. Therefore, the lab frame tip position $\ptip^\a$ is given by:
\[
 \ptip^\a = \pfil^\a + \R\a\A (\rPhi) \dfil^\A. 
\]
% \esub

%Eqs. \eq{labcomov} show the freedom in picking a reference frame: one part of the lab frame motion $(\ptip^\a, \rphi)$ is attributed to the frame motion $(\pfil^\a, \rPhi)$,
%leaving a 'residual motion' $(\dfil^\A, \rPhi)$ relative to the moving frame. 

Differentiating %Eqs. \eq{labcomov} 
with respect to time delivers
\alnn{
 \dd_\t \ptip^\a &= \dd_\t \pfil^\a + \R\a\A (\rPhi) \mOmgo \dd_\mPsi \dfil^\A - \romgo \lcsf\a\b \R\b\A(\rPhi) \dfil^\A. 
}

\subsection{Symmetry breaking}

The Euclidean plane $\Real^2$ is invariant under the Euclidean group $\SE2$ of translations in $\x$ and $\y$ and rotations. However, the spiral wave solutions are not invariant under $\SE2$, since shifting or rotating a solution will yield a solution to the RDE \eq{RDE} that is distinct from the original one. So, at least $3$ symmetries are broken by the spiral wave solution. 

As in the case of 1D pulses, one may introduce a moving frame of reference, which in this case is also rotating; see below for different options. 
If in that frame, the spiral wave is stationary, one has $\NBS=3$ and the spiral wave is rigidly rotating; see Par. \ref{sec:rigidspiral}. Else, the time symmetry is also broken and $\NBS=4$. Here, we consider only the case where the spiral wave solution in the co-moving frame of reference is periodic in time. Returning to the lab frame of reference, this case represents a meandering spiral wave, either in the non-resonant (see Par. \ref{sec:meanderspiral}) or resonant case (see Par. \ref{sec:resonant}). 

\subsection{Rigidly rotating spiral waves \label{sec:rigidspiral}}
%Spiral waves are persistent solutions to Eq. \eq{RDE} that appear as rotating rather than translating over time. Their dynamics is commonly characterised by tracing the spiral wave tip, \ie\ a point that can be uniquely determined at each instance of spiral wave rotation. 
%A rigidly rotating spiral has three degrees of freedom: one should specify its center or rotation and the rotation phase $\rPhi$. However, spiral wave motion is known to be determined by processes taking place at the spiral wave tip. 

\begin{figure}
\raisebox{3cm}{a)} \includegraphics[width=0.19\textwidth]{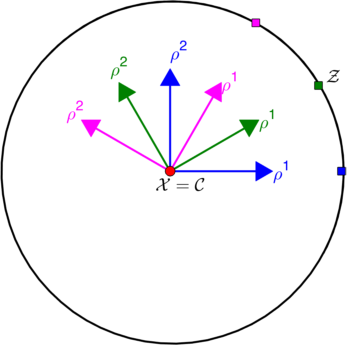}
\raisebox{3cm}{b)} \includegraphics[width=0.22\textwidth]{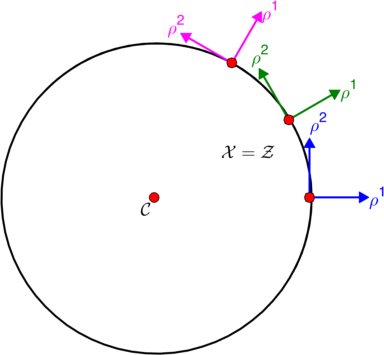}
\caption{Different coordinate systems suitable to analyse and circular-core spiral wave. a) Center frame: the origin is taken at the center of the circular tip trajectory. b) Tip frame: the origin is taken at the current position of the spiral wave tip, and the $\xp^2$ axis is tangent to the tip trajectory. \label{fig:frames_circ} }
\end{figure}

%Even for this simplest case, we have multiple options to choose a co-moving frame of reference, see Fig. \ref{fig:frames_circ}. 

When going to a frame of reference rotating at $\romgo=\K\somg$, (\ie\ the spiral wave frequency), the solution becomes periodic: 
\aleq{u0rigidspiral}{
 \UU(\x^\a,\t) &= \uuo(\xp^\A), \\
 \dd_\t \X^\A &= \vo\A, \\
 \dd_\t \rPhi &= \romgo 
}
 
with $\vo\A$ and $\romgo$ constant. We distinguish two frames of reference here: 

\subsubsection{Center frame}

The simplest coordinate system is found when taking its origin in the middle of the circular tip trajectory, as shown in Fig. \ref{fig:frames_circ}(a), so that
\alnn{
  \vo\A = 0. 
}
 
 This frame was used in the vast majority of previous works on spiral wave theory \cite{Keener:1988, Biktashev:1994, Verschelde:2007, Dierckx:2009, Biktashev:2010, Dierckx:2012, Dierckx:2013, Dierckx:2013virtual, Li:2014}. 
This approach, however, is not optimal for describing the interaction of the spiral wave with localised heterogeneities, as dynamics is most influenced by stimuli close to the spiral wave tip rather than rotation centre, and the distance between the two can be significant. 

\subsubsection{Tip frame}
Therefore, we introduce another frame of reference, with the origin of coordinates $\xp^\A$ at the spiral wave tip, 
$\cfil = \ctip$
as shown in Fig. \ref{fig:frames_circ}b:
\alnn{
   \vo\A \neq 0. 
}
 
 In this `tip frame', the velocity of the spiral wave tip $\vo\A$ becomes constant. One can even choose the orientation of the frame such that $\vo1=0$, $\vo2=\romgo\rt$ with $\rt$ the radius of the tip trajectory. 

The tip frame closely relates to the 
kinematic approach of Zykov et al. \cite{Zykov:book} and is particularly suited to describe large-core spiral waves. 

Finally, note that the freedom in the choice of the definition of the tip may be exploited by taking
${\uc\j}=\u_\j(\ccent)$, ${\uc\k}=\u_\k(\ccent)$,
so that $\ctip=\ccent$; then we have
$\vo\A=0$ and recover the center frame as a special case. 

\subsection{Meandering spiral waves \label{sec:meanderspiral}}

The case of biperiodic meander studied in this work is characterized by the fact that the spiral wave solution can be made time-periodic in a moving frame of reference. As in the case of periodically modulated waves, we introduce a temporal phase $\mPsi$, which in the absence of perturbations increases from $0$ to $2\pi$ over the time interval $\To$: $\mPsi(\t) = \mOmgo\t$, $\mOmgo = 2\pi/\To > 0$. Without loss of generality, we may take $\Psi(0)=0$. In the context of spiral waves, we refer to $\mPsi$ as the meander phase, or temporal phase, as opposed to the rotation phase (angle) $\rPhi$.

A meandering spiral solution is thus of the form:
\algrp{u0meanderspiral}{
 \UU(\x^\a,\t) &=  \uuo(\xp^\A, \mPsi(\t)),     \eqlabel{u0meanderspiral-UU}\\
 \dd_\t \X^\A &= \vo\A(\mPsi),                  \eqlabel{u0meanderspiral-dtX}\\
 \dd_\t \rPhi &= \romgo(\mPsi),                 \eqlabel{u0meanderspiral-dtphi}\\ 
 \dd_\t \mPsi &= \mOmgo.                        \eqlabel{u0meanderspiral-dtpsi}
}
 
with $\uuo$ $2\pi$-periodic in $\mPsi$.

An overview of reference frames discussed below is given in Tab. \ref{tab:2D}. 

The Euclidean transformation mapping $\uu(\x,\y,\t)$ to $\uu(\x,\y,\t+\To)$ can, without loss of generality, be always thought of as either a pure rotation or a pure translation, since the composition of a rotation and a translation can always be written as a single rotation in 2D. Below we first treat the `pure rotation' case, corresponding to non-resonant meander; the second case (pure translation), corresponding to resonant meander will be discussed in Sec. \ref{sec:resonant}. 

\begin{figure*} \centering
%\begin{tabular}{cc}
%Flower-like meander &
%Linear-core meander
%\end{tabular}\\
\raisebox{4cm}{a)} \includegraphics[width=0.3\textwidth]{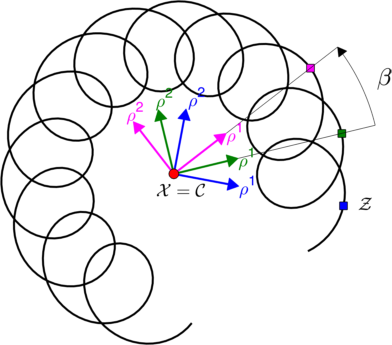}
\raisebox{4cm}{b)} 
\includegraphics[width=0.3\textwidth]
{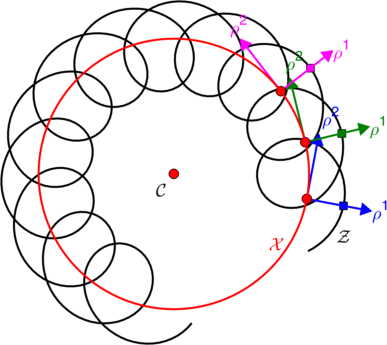}
\raisebox{4cm}{c)} 
\includegraphics[width=0.3\textwidth]{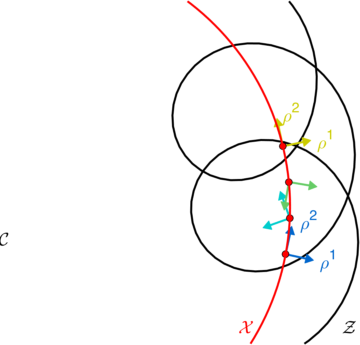}
\\
\raisebox{4cm}{d)} \includegraphics[width=0.3\textwidth]{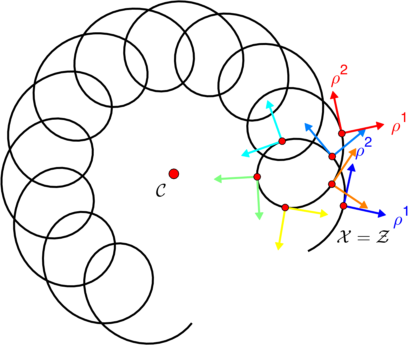}
\raisebox{4cm}{e)} 
\includegraphics[width=0.3\textwidth]
{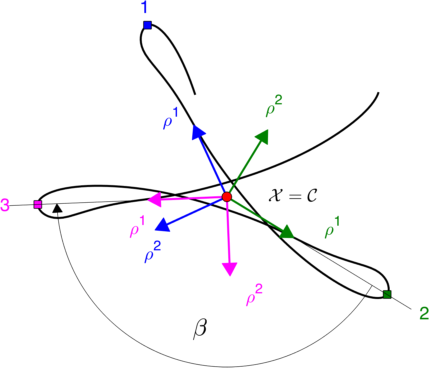}
\raisebox{4cm}{f)} 
\includegraphics[width=0.3\textwidth]{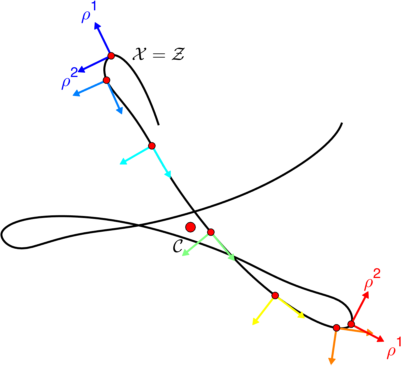}
\caption[]{Different coordinate systems suitable to analyse and predict the motion of a meandering spiral wave. Panels a-d show cycloidal meander in Barkley's model with parameters as in Fig. \ref{fig:patterns}d; panels e-f illustrate linear-core meander in the FK-Guinea Pig cardiac model. Panels (a,e) show the `center frame', where the origin is taken at the center of meander flower, with the frame rotating at the constant rate $\romgo = \alp/\To$, where $\alp$ is the angle between consecutively visited petals. Panels (b,c) illustrate frames tailored describe the finite-core, which may either minimally rotate (over an angle $\alpo$ (see b) or co-rotate with the spiral, \ie\  rotate over $\alps = \alpo + 2\pi \K$ where $\K=\pm 1$ labels chirality. Panels (d,f) show a tip frame of reference, with the origin $\cfil$ of the coordinate system at the current tip position and the $\xp^2$-axis tangent to the tip trajectory. The rotation rate of the tip frame is  $\romgo(\mPsi)$, non-constant but periodic. 
\label{fig:frames_meander}}
\end{figure*}

\subsubsection{Definition of the angle $\alp$ between petals}

In the case of non-resonant meander, the spiral pattern repeats itself after time $\To$, but rotated around an angle $\alp$ around a point $\ccent$ that will become the center of the spiral tip trajectory (`meander flower'); see Fig. \ref{fig:frames_meander}. 

Since we only impose that the solution be periodic, $\alp$ can be chosen up to an integer multiple of $2\pi$. Two values can play a special role, however. First, we can define this angle by following the movement of the spiral. If we attach a reference direction to the spiral wave tip (\eg\ $\vec{\nabla}\u_1$ evaluated at the tip), this vector will turn over the angle $\alps$ during one spiral period, and for a faraway observer, the average spiral wave frequency (expressed as a positive number) will be $\somg = \K \alps /\T$, where $\K = \pm 1$ denotes chirality: $\K=1$ for counterclockwise and $\K=-1$ for clockwise rotation of both the spiral wave and the frame of reference. 

Secondly, we may choose the $\alp$ as the element of the set $\alps + 2\pi\Zahlen$ that has the minimal absolute value, and call this minimal value $\alpo$.
For the cases studied here, we find
\alnn{
 \alps = \alpo + 2\pi\K % \eqlabel{betas}
}
 
with $0 < |\alpo| < \pi$. 

\subsubsection{Center frame}

We pick a frame rotating at constant frequency $\romgo = \alpo/\To$ around $\ccent$, \ie\ we take $\cfil = \ccent$. In this `center frame of reference', the meandering spiral solution becomes periodic. The solution is given by \eq{u0meanderspiral}, with 
\alnn{
 \vo\A = 0, \qquad \romgo = \alpo/\T. 
}

This set of equations describes the spatial position the spiral wave in terms of the position of the center of the meander flower, and was used in \cite{Dierckx:2017PRL}. Since we chose $\cfil = \ccent$, Eq. \eq{u0meanderspiral-dtX} expresses that the center of the meander flower does not move in the absence of perturbations to the RDE \eq{RDE}. 

It is also possible to relax the convention to take the period $\To$ as the minimal time interval in which the solution is periodic modulo Euclidean group actions. E.g. when linear cores rotate over $\alp\approx 180^\circ$ \cite{Fenton:1998}, one could define the period to span two such cycles, bringing $\alp \approx 0^\circ$ instead. This formalism can then be used to study phase-locking to constant external fields, generalising the results in \cite{Dierckx:2017PRL}. 

Another theoretical possibility is to use the same setting as described by system
\eq{u0meanderspiral}, but with the choice $\alp=\alps$ instead of $\alpo$. This choice would ensure that the spiral wave solution, considered in the rotating frame, oscillates but not rotates and remains approximately stationary far from the centre. Neither of these two theoretical alternatives is used later in this paper, so we mention them only for completeness.

\subsubsection{Minimally rotating finite-core frame}
It can be advantageous to allow the origin of the coordinate system to better follow the spiral wave tip. One possible choice is shown in Fig. \ref{fig:frames_meander}b for reaction kinetics with cycloidal meander. We let the new origin move at constant angular velocity on a circular trajectory that approximates the exact tip trajectory, and refer to this situation as the `minimally rotating finite-core frame':
\alnn{
  \dd_\tp \vo\A =0, \qquad \romgo = \alpo/\To.  
}

\subsubsection{Co-rotating finite-core frame}
We can let the minimally rotating finite core frame rotate around its axis, at a rate $\K \mOmgo$, to better follow the spiral wave rotation, bringing the absolute rotation rate to $\alps$; see Fig. \ref{fig:frames_meander}c. We will refer to this case as the `co-rotating finite-core frame'. It is again described by Eqs. \eq{u0meanderspiral}, with 
\al{
  \dd_\tp \vo\A =0, \qquad \romgo = \alps/\To.  
}
 
In this frame, the spiral solution $\uuo(\xp^\A, \mPsi)$ will not rotate if $\mPsi$ is increased from $0$ to $2\pi$.  

\subsubsection{Tip frame}

Lastly, we let the origin $\cfil$ of the co-moving frame of reference coincide with the spiral wave tip position, \ie\ $\cfil = \ctip$, $\pfil^a(\t) \equiv \ptip^a(\t)$; see Fig. \ref{fig:frames_meander}c and f. Then $\vo\A$ and $\romgo$ become non-constant, $2\pi$-periodic functions of $\mPsi$. 

After one meander period of duration $\To$, the frame will have turned over an angle
\alnn{
  \alps 
  = \int_0^{\T} \romgo(\mPsi(\t))\,\d\t 
  = \frac{1}{\mOmgo} \int_0^{2\pi} \romgo(\mPsi) \,\d\mPsi  
  =  \bar{\romgo} \To. 
}

If the tip trajectory is smooth, we can without loss of generality suppose that $\vo1 = 0$, $\vo2 = \vo{}(\mPsi)$. This can always be achieved by exploiting the freedom in the definition of frame orientation angle $\rPhi$. 

\subsubsection{Comparison with the classical theory of meander}

In the classical theory of rigidly rotating spiral waves, the critical eigenmodes for translation have eigenvalues $\pm\iu\romgo_1$, where $\romgo_1$ is the rotation frequency of the rigidly rotating spiral wave. If due to a parameter change in the medium, another complex eigenvalue pair crosses the imaginary axis at $\pm\iu\romgo_2$, we have a Hopf bifurcation in the quotient system (rotating frame). If this bifurcation is supercritical, this results in an epicycloidal tip trajectory as shown in Fig. \ref{fig:frames_meander}(a--c).  In Barkley's notation \cite{Barkley:1994}, it is seen that beyond the Hopf bifurcation the absolute tip velocity will oscillate at $\romgo_2$. Hence, we conclude that Barkley's $\romgo_2 $ equals $\mOmgo$ used throughout this work. Secondly, in the circular-core regime just before the Hopf bifurcation, the spiral's rotation frequency is $\romgo_1$, corresponding to $\somg$ in our notation. We conclude that in the tip frame (where $\alp = \alps$), $\romgo = \romgo_1 $ and in the minimally rotating frames, (where $\alp = \alpo$), $\romgo= \romgo_1 - \K \mOmgo$. 

The case of resonant meander happens in the classical theory when $\romgo_2 \rightarrow \romgo_1$. In our description, this happens when $\romgo \rightarrow 0$ in the center or minimally co-rotating frame, or when $\romgo \rightarrow \mOmgo$ in the tip frame or fully co-rotating frame.

\subsection{Resonant meander \label{sec:resonant}}
We now consider the case where the meandering spiral returns after the temporal period $\To$ to the same state, but translated over a vector $\dist^\a = \avg{\vo\a} \To$;
see Fig. \ref{fig:patterns}(f). Different authors call this sort of solutions 
cycloidal motion~\cite{Zykov-1986},  
$0^\circ$ isogon contours~\cite{Winfree:1991},
modulated travelling waves~\cite{Barkley:1994,Wulff:1996},
resonant linear motion~\cite{Golubitsky:1997} or
linear drift~\cite{Ashwin-etal-2001}.
We shall refer to it here as resonant meander.

Note that during dynamics under external perturbations, the direction of spontaneous drift may change, such that a rotation phase needs to be introduced nevertheless. This makes this case distinct from the 1D modulated traveling wave from Par. \ref{sec:MTW}. 

The solution is still given by \eq{u0meanderspiral}, with $\vo\A$ and $\romgo$ depending on the chosen reference frame: 

\subsubsection{Minimally rotating co-moving frame}
\begin{figure} \centering
\raisebox{4cm}{a)} \includegraphics[width=0.2\textwidth]{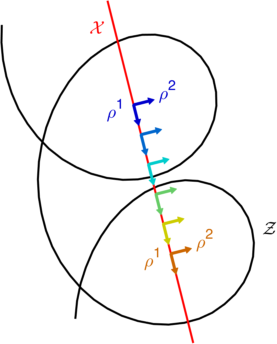}
\raisebox{4cm}{b)} \includegraphics[width=0.2\textwidth]{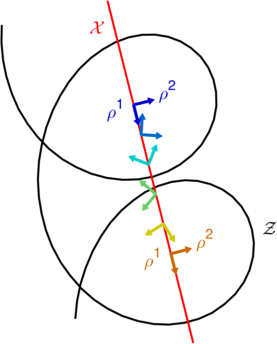} \\
\raisebox{4cm}{c)} 
\includegraphics[width=0.2\textwidth]{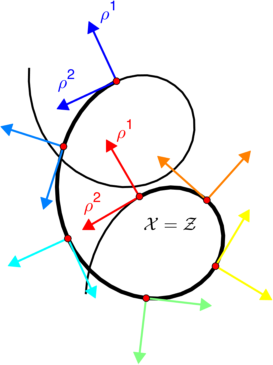}
\caption{Reference frames for resonant meander. a) The `minimally rotating co-moving frame' is rigidly translating such that the solution becomes periodic. During one period (\ie\ between the frames shown), the frame rotates over an angle $2\pi\K$.  
b) The `co-rotating co-moving frame' is obtained from the average frame by letting the moving coordinates rotate at constant frequency $\K \mOmgo = 2\pi\K/\To$.  c) Tip-based frame of reference, with the origin of the coordinate system at the current tip position and the $\xp^2$-axis tangent to the tip trajectory. Rotation rate of the frame is $\romgo(\t)$, non-constant but periodic. 
\label{fig:frames_resonantspiral}}
\end{figure}

First, we let the origin of our coordinate system move on a straight line with velocity $\vo1,\vo2$, without frame rotation, parallel to the line along which the resonant spiral travels over time.  This situation is sketched in Fig. \ref{fig:frames_resonantspiral}a. Then, we find:
\alnn{
  \dd_\tp \vo\A = 0, \qquad \dd_\tp \rPhi = 0. 
}

Since the frame is not rotating, the function $\uuo(\xp^\A,\mPsi)$ shows a spiral that rotates a full turn when $\mPsi$ is raised from $0$ to $2\pi$.

\subsubsection{Co-rotating, co-moving frame}
We can also let the previous frame rotate at $\romgo = \K \mOmgo$, while its origin moves on a straight line at constant speed parallel to the displacement direction of the resonant spiral. We again obtain Eqs. \eq{u0meanderspiral}, with 
\alnn{
  \vo\A = \R\A\a(\rPhi) \vo\a, \qquad \dd_\tp \rPhi = \K \mOmgo. 
}
 
and $\vo\a$ constant. The spiral wave profile $\uuo(\xp^\A,0,\mPsi)$ does not rotate a full turn when $\mPsi$ is raised from $0$ to $2\pi$; it only deforms (`breathes'), see \eg~\cite{Foulkes:2010}.

\subsubsection{Tip frame}
As in the non-resonant case, it is also possible to find a tip-based frame of reference: 
\alnn{
  \rPhi = \rphi, \qquad \pfil^\a = \ptip^\a
}
 
Note that during the time interval $\T$, the angle $\rPhi$ needs to turn over $\pm 2\pi$, whence
\alnn{
  \avg{\romgo} = \K \mOmgo. 
}

Here, one can see that resonant meander is indeed a special case of Eqs. \eq{u0meanderspiral}, in which the average value of ${\romgo}$ tends to $\K \mOmgo$. In the tip frame, $\uuo$ represents a spiral that performs no net rotation when $\mPsi$ is raised from $0$ to $2\pi$, it merely deforms around the same orientation.

\subsection{General case}

The foregoing patterns can all be captured by noting that each broken continuous symmetry delivers a collective coordinate $\X^\M$:
\alnn{
  \x \rightarrow \X^1, \qquad
  \y \rightarrow \X^2, \qquad
  \polang \rightarrow \rPhi, \qquad
  \t \rightarrow \mPsi.
}
 
We shall denote this as $\x^\M\rightarrow\X^\M$; there is one pair for every broken continuous symmetry. We also convene that 
\alnn{
  \{ \X^\M \} &= \{ \X^\A, \rPhi, \mPsi\}, \\
  \{ \X^\m \} &= \{ \X^\a, \rPhi, \mPsi\}.
}
 
The use of indices $\M$ vs. $\m$ thus distinguishes between the moving and lab frames of reference, as did $\A$ vs. $\a$
 for the translational modes only. Note that $\X^\a$ is tip position in the lab frame, while $\x^\a$ is a lab frame Cartesian coordinate that can label any point. 

The parameters of the solution $\x^\M $ are chosen from $\{\x,\y,\polang,\t\}$ and $\X^\M$ are the collective coordinates. Then we find that all aforementioned sets of equations \eq{u0wave}, \eq{u0deformingwave}, 
\eq{u0rigidspiral}, 
and \eq{u0meanderspiral}
that capture the evolution of particular reaction-diffusion patterns can be written as
\aleq{u0general}{
 \UU(\x^\a, \t) &=  \uuo(\xp^\A, \mPsi),  \\
 \dd_\t \X^\M &= \vo\M(\mPsi) 
}

given a time-dependent spatial coordinate transformation
\alnn{
 \xp^\A(\x^\a, \t) &= \gg(\x^\a, \X^\M(\t)) 
}
that has as many degrees of freedom as broken Galilean symmetries by the solution. 

Note that the representation \eq{u0general} is not unique. E.g. for a rigidly rotating spiral wave, one could choose $\UU(\x^\a, \t) = \uuo(\xp^\A, \mPsi)$ that is periodic in its third argument $\mPsi \equiv \rphi$ and $\xp^\A$ identical to the lab frame coordinates $\x^\a$. However, one can also choose to let $\uuo$ in the right-hand side of Eq. \eq{u0general} only depend on $\xp^\A$ and $\mPsi$.

For the actual calculations performed below, we work with Eqs. \eq{u0meanderspiral} for a meandering spiral in a tip-based frame of reference, since they contain all other patterns and reference frames as a special case, \ie\ Eqs. \eq{u0wave}, \eq{u0deformingwave}, \eq{u0rigidspiral}and \eq{u0meanderspiral}. 
Those can be recovered from Eq.  \eq{u0meanderspiral} by considering $\uuo$ independent of a collective coordinate. Thereafter, the evolution equation for that coordinate becomes irrelevant and can be dropped. 
E.g. the theory of rigidly rotating spirals follows from the meander case by taking $\uuo$ independent of $\mPsi$ and leaving out the evolution equation for $\mPsi$. From this case, the 1D traveling wave case follows by considering $\uuo$ to be also independent of $\rPhi$ and $\xp^2$, such that Eq. \eq{u0wave} follows. 

\section{Properties of the linearized problem \label{sec:linear}}

\subsection{Right-hand zero modes}

Since the original RD Eq. \eq{RDE} is invariant under spatial, rotational and temporal shifts, one has that, when a solution is rotated, or shifted in space or time, it is still a solution to Eq. \eq{RDE}. This can be made explicit by substituting $ \uu(\x+\shift, \y, \t)$ in Eq. \eq{RDE}, and similar for other coordinates, yielding in first order in $\shift$ that
\al{
  \linop \  \df{\UU}{\x^\M} = \bzero. \eqlabel{zeromode}
}
 where 
\al{
  \linop  = \HP (\dd_\x^2 + \dd_\y^2) + \bF'(\UU) - \dd_\t.  \eqlabel{llab}
}
As expected, shifted values of the solution are right-hand zero modes of the linearized operator $\linop$ associated to the RDE, \textit {in the lab frame.}

In several previous works, a moving frame was chosen in which the solution was either stationary \cite{Keener:1986, Biktashev:1994, Verschelde:2007, Dierckx:2009, Dierckx:2012} or periodic \cite{Dierckx:2017PRL}. This comes down to expressing the derivative in Eq. \eq{zeromode} in the moving frame using the chain rule. From Eqs. \eq{u0meanderspiral} then follows 
\aleq{dtu0}{
  \dd_\t \UU &= \dd_\A \uuo \dd_\t \xp^\A + \dd_\mPsi \uuo \dd_\t \mPsi \\
  & = \dd_\A \uuo [\romgo \R\A\a(\rPhi)  \lcsf\a\b  (\x^\b - \X^\b(\t)) - \R\A\a(\rPhi) \vo\a] \\
  & = \mOmgo \dd_\mPsi \uuo - \romgo \dd_\polang \uuo - \vo\A \dd_\A \uuo. 
}

Here, $\polang$ is the polar angle around the origin of the chosen moving reference frame. Thus, $\uuo$ obeys
\al{\eqlabel{RDEcomov}
  \HP \Lap \uuo + \bF(\uuo) + \romgo \dd_\polang \uuo + \vo\A \dd_\A \uuo  - \mOmgo \dd_\mPsi \uuo = \mathbf{0}. 
}
 
Here, $\romgo$ and $\vo\A$ may depend on $\mPsi$ (which is the case in the tip frame of a meandering spiral wave). The linear operator associated to Eq. \eq{RDEcomov} is
\alnn{
  \CL = \HP \Lap  + \bF'(\uuo) + \romgo \dd_\polang + \vo\A \dd_\A  - \mOmgo \dd_\mPsi
}

Since the original RDE \eq{RDE} is invariant in space, one has that in an infinite medium, if $\UU(\x^\a,\t)$ is a solution, so should $\UU(\x^\a+\shift^\a, \t)$ for any constant displacement $\shift^\a$. We find in one spatial dimension that
\alnn{
 \CL \dd_{\xp} \uuo = \bzero. 
}
In two spatial dimensions, we remark that $\dd_\a \UU = \dd_\A \uuo \, \R\A\a(\rPhi)$, 
such that we find in first order in $\shift^\a$: 
\al{\eqlabel{eigenmodes2D}
  \CL \dd_\A \uuo = \romgo \lcsb\A\B  \dd_\B \uuo. 
}
 
By taking complex combinations $\VV_\pm = -(1/2)(\dd_\x \uuo \pm \iu\dd_\y \uuo)$, one obtains true eigenmodes of $\CL$, as $\CL \VV_{\pm} = \pm \iu \romgo \VV_\pm$ \cite{Biktasheva:2009}.  Note that in some of the chosen reference frames, $\romgo$ may depend on $\mPsi$, and the `eigenmodes' of the system are $2\pi$-periodic functions of $\mPsi$. 

For 2D patterns, we can state that a solution that is rotated around an angle $\shift$ around the current origin (\ie\ spiral tip position or meander centre position) will still be a solution. 
With 
$\dd_\polang = \lcsb\A\B \xp^\A \dd_\B$, 
one finds that, if 
$\UU(\x^\a, \t) + \shift \dd_\polang \UU(\x^\a,\t)$ is a solution, then
\alnn{
  \CL \dd_\polang \uuo = \bzero. 
}
 
We will denote $- \dd_\polang \uuo $ as $\VV_\rPhi$ and $-\dd_\A \uuo $ as $\VV_\A$. 

Thirdly, expressing that a time-shifted solution $\UU(\x^\a, \t+\shift)$ also solves \eq{RDE} yields
\alnn{
  \CL \dd_\t \uuo = \mathbf{0}. 
}
 
where $\dd_\t \uuo$, also denoted $\VV_\mPsi$,
 is given by Eq. \eq{dtu0}. 
%\esub
 
In the case of meandering spirals, $\VV_\rPhi$ is linearly independent of $\VV_\mPsi$, since otherwise a shift in time would be equivalent to simple rotation, and we would find ourselves in the non-meandering case. 

  To summarise, equation~\eq{llab} defines the set of $\NBS$ zero modes for the linearized operator $\linop$ in the laboratory frame; in the comoving frame, according to \eq{eigenmodes2D}, this produces a set of $\NBS$ eigenvalues 
\alnn{
  \HL \VV_\pm = \pm \iu \romgo \VV_\pm, \qquad
  \HL \VV_\rPhi = \bzero , \qquad 
  \HL \VV_\mPsi  = \bzero. 
}
 
In quantum field theories, bosons appearing due to spontaneous breakdown of continuous symmetries are called Goldstone bosons; extending the analogy to the classical nonlinear field, the eigenfunctions corresponding to breakdown of continuous modes are sometimes referred to as `Goldstone modes'. 

% Need to list explicitly the full set of evalues and efuncs in one place. Currently it is scattered or hinted upon.
% reply: see Eq. 38 single equation number containing all cases. 

\subsection{Adjoint problem}

Let us associate %to $\linop$ the operator
to $\CL$ the operator
\alnn{
  \CLd = \HP^T \Lap  + \bF'^T(\uuo) - \romgo \dd_\polang - \vo\A \dd_\A  + \mOmgo \dd_\mPsi . 
}
Note that $\uuo$ is $2\pi$-periodic in $\mPsi$, and in the space of functions $2\pi$-periodic in $\mPsi$,
$\CLd$ is the adjoint operator to $\CL$, 
in the sense that
\alnn{
  \bbraket{\CL \ffield}{\gfield} = \bbraket{\ffield}{ \CLd \gfield}
}
where
\algrp{inner}{
  \bbraket{\ffield}{\gfield} &= \frac{1}{2\pi} \int_0^{2\pi} \braket{ \ffield}{\gfield} \,\d\mPsi,  \eqlabel{innera} \\
  \braket{\ffield}{\gfield} &= \int_{\Real^\Ns} \ffield^H \gfield \,\d^\Ns\x ,                      \eqlabel{innerb}
}

where $\Ns=1$ for 1D waves and $\Ns=2$ for 2D wave patterns. 
For non-deforming solutions, we note that $\uuo$ is independent of $\mPsi$, so the inner products in Eqs. \eq{innera}
and \eq{innerb} coincide, for any $\ffield$ and $\gfield$ defined by $\uuo$.

Given that $\CLd$ is the adjoint to $\CL$, we assume that it also has $\NBS$ critical eigenmodes $\WW^\M$ that are $2\pi$-periodic in $\mPsi$:
\al{
  \CLd \WW^\pm = \mp \iu \romgo \WW^\pm, \qquad 
  \CLd \WW^\rPhi = \bzero, \qquad
  \CLd \WW^\mPsi = \bzero.  \label{defW}
}

These `adjoint critical modes' are known as sensitivity functions or response functions (RFs) \cite{Biktasheva:2003}, as will be explained in Sec. \ref{sec:drift}. 

\subsection{Instant orthogonality of left and right critical modes}

The set of critical adjoint modes can be normalised as
\alnn{
  \bbraket{\WW^\M}{\VV_\N} = \kron\M\N. 
} 
 
Moreover, the orthogonality of critical and adjoint critical eigenmodes holds instantaneously~\cite{Foulkes:2009, Marcotte:2016, Dierckx:2017PRL}:
\al{
  \braket{\WW^\M}{\VV_\N} = \kron\M\N, \eqlabel{meanderlemma}
} 
Here, we show where the proof in \cite{Dierckx:2017PRL} needs to be adapted to accommodate for non-constant rotation rates $\romgo(\mPsi)$ and the case of resonance ($|\romgo| \rightarrow \mOmgo$). 

Let us suppose that $\CLd \WW^\M = \conj{\la_\M} \WW^\M$ and $\CL \VV_\N = \la_\N \VV_\N$. 
Let us define the operators
\al{
 \HL =  \CL + \mOmgo \dd_\mPsi, 
 \qquad
  \HLd =  \CLd - \mOmgo \dd_\mPsi .
}
We note that
$\HLd$ is adjoint to $\HL$ with respect to the inner product $\braket{\cdot}{\cdot}$, which follows from integration by parts and the tempered nature of the adjoint eigenmodes $\WW^\M$.  
Denoting $\braket{\WW^\M}{\VV_\N} $ as $\I\M\N$, it follows that
\aleq{dpsiI}{
   \mOmgo \dd_\mPsi & \I\M\N \\
  =& \braket{(\HL\adj- \CL\adj)\WW^\M}{\VV_\N} +  \braket{\WW^\M}{(\CL - \HL)\VV_\N} \\
  =& (\la_\N - \la_\M)  \I\M\N
     -  \braket{\HL^\dagger \WW^\M}{\VV_\N}
     +  \braket{\WW^\M}{\HL\VV_\N} \\
  =& (\la_\N - \la_\M)  \I\M\N. 
}
Hence 
\al{
  \I\M\N(\mPsi) = \I\M\N(0) \exp\left( \frac{ \int_0^\mPsi( \la_\N(\z) - \la_\M(\z)) \,\d\z)}{\mOmgo}\right). \eqlabel{mesol}
}
Now, if we have $\la_\M = \la_\N$, then $\I\M\N$ is constant. Then, setting $\bbraket{\WW^\M}{\VV_\N} = \kron\M\N$ already yields \eq{meanderlemma}. Else, if $\Re(\la_\M) \neq \Re(\la_\N)$, and $\I\M\N(0)\ne0$, then $\abs{\I\M\N(\mPsi)}$ grows or decays exponentially in time, which cannot happen since it should be $2\pi$-periodic. Therefore, $\I\M\N(\mPsi)=0$ in this case.
Finally, it is possible that $\Re(\la_\M) = \Re(\la_\N)$ but $\Im(\la_\M) \ne \Im(\la_\N)$, \eg\ when considering the inner product between critical eigenmodes. In that case, $\la_\M-\la_\N=\iu\romgo(\mPsi)\mint$ with $\mint \in\{0, \pm 1, \pm 2 \}$. Then, 
 \al{
\frac{\int_0^{2\pi}( \la_\N(\z) - \la_\M(\z)) \,\d\z) }{\mOmgo} = \iu\mint\alp .
}
If not in the resonant case, $\alp \mod \pi \neq 0$, and the exponential factor in Eq. \eq{mesol} cannot be periodic, whence $\I\M\N=0$. 
%
%(Hans): after correction for 2pi case it comes down to exp(\iu \pi)  = -1. So better, but no proof for the meander lemma this way. 
In the case of resonance, we are free to choose a non-rotating frame, where $\romgo=0$, such that \eq{meanderlemma} still holds.   \hfill\ensuremath{\square}

In the laboratory frame of reference, the critical modes are true zero modes, whence all $\la_\m = 0$, and the preservation of the inner product immediately follows from Eq. \eq{dpsiI}. 

\section{Spatiotemporal drift of patterns under a small perturbation \label{sec:drift}}

\subsection{Derivation of the drift equations}

Using the ingredients defined above, it is possible to predict how a stable reaction-diffusion pattern (\eg\ a plane wave or a spiral wave) reacts to a small perturbation. We mainly follow the derivation for rigidly rotating spiral waves \cite{Biktashev:1995b} but extend it to the case of meander. In comparison to~\cite{Dierckx:2017PRL}, we offer more flexibility in the frame of reference, such that also meandering spirals close to resonance can be treated. 

We start from the perturbed RDE:
\al{
 \dd_\t \uu = \HP (\dd^2_\x + \dd^2_\y) \uu + \bF(\uu) + \hh(\x,\y,\t) \eqlabel{RDEpert}
}
where $\hh = \OO(\heta)$ 
is a small perturbation.
A more generic form of perturbation can include dependence on the solution itself, \ie\ $\hh(\x,\y,\t,\uu,\nabla)$ which however is reduced to \eq{RDEpert} when the perturbation is evaluated at the unperturbed solution, \ie\ $\hh(\x,\y,\t,\uuo(\x,\y,\t),\nabla)$.

If the initial state is close to a stable solution (\ie\ travelling or spiral wave), the net effect of $\hh$ will be to cause a spatiotemporal drift of that pattern, which can be inferred from the collective coordinates. As before, we will present the result from the most general case of a meandering spiral wave (Eqs. \eq{u0meanderspiral}), from which all other cases can be inferred by eliminating some of the collective coordinates. 

Thus, we approximate the solution as
\al{
 \uu(\x^\a, \t) =  \uuo(\xp^\A, \mPsi) + \tuu(\x^\a,\t), \eqlabel{udec}
}

where $\tuu = \OO(\heta)$. The coordinate transformation is now given by
\alnn{
  \xp^\A &= \R\A\a(\rPhi(\t)) (\x^\a - \X^\a(\t)), \\
  \mOmgo \tp &= \mPsi(\t). 
}
in which the collective coordinates' temporal evolution is perturbed by yet unknown drift terms $\pv\M$, which are also $\OO(\heta)$:
\alnn{
 \dd_\t \X^\a &= \R\A\a(\rPhi) (\vo\A + \pv\A), \\
 \dd_\t \rPhi &= \romgo(\mPsi) + \pv\rPhi, \\ 
 \dd_\t \mPsi &= \mOmgo + \pv\mPsi.
}

We can make the decomposition \eq{udec} unique by imposing that
\al{
  \braket{\WW^\M}{\tuu} = 0. \eqlabel{gaugeu}
}
This is possible by shifting the solution. E.g. if one approximates at a given instance of time the solution $\uuo(\xp^1, \xp^2, \mPsi)$ as $\uuo(\xp^1 + \shift, \xp^2, \mPsi) + \shift \dd_1 \uuo(\xp^1, \xp^2, \mPsi)$, 
then $\braket{\WW^{1}}{\tuu}  = \shift \neq 0$, and the proposed solution can be better shifted to match the true solution, \ie\ until Eq. \eq{gaugeu} holds.  A second interpretation of Eq. \eq{gaugeu} is that the deviation vector $\uu$ should be orthogonal to the inertial manifold~\cite{Biktashev:1995b}.

Next, plugging the ansatz \eq{udec} into the perturbed RDE \eq{RDEpert} delivers: 
\alnn{
  \dd_\tp \tuu - \CL \tuu + \pv\M \VV_\M = \hh + \OO(\heta^2). 
}
 
Note that by including the minus sign in the definition of the GMs, \ie\ $\VV_\rPhi = - \dd_\polang \uuo$, $\VV_\A = - \dd_\A \uuo$, $\VV_\mPsi = + \dd_\mPsi \uuo$, a plus sign appears in front of $\pv\M \VV_\M$. 
Projection onto the response function $\WW^\M$ delivers, due to the meander lemma \eq{meanderlemma}: 
\alnn{
  \braket{\WW^\M}{\dd_\tp - \HL \tuu} - \pv\M =  \braket{\WW^\M}{\hh} + \OO(\heta^2). 
}
The first term vanishes, due to the gauge condition \eq{gaugeu} and the fact that $\HL$ is adjoint to $\HL$ with respect to $\braket{\cdot}{\cdot}$:
\alnn{
  \braket{\WW^\M}{(\HL- \dd_\tp) \tuu} 
  &= \braket{(\HL^\dagger + \dd_\tp ) \WW^\M}{\tuu} \\ 
  &= \braket{(\HL^\dagger + \mOmgo \dd_\mPsi) \WW^\M}{\tuu} \\
  &= \braket{\CL \WW^\M}{\tuu} \\
  &= \la_\M \braket{\WW^\M}{\tuu}  = 0. 
}
 
Hence, we find the simple result: 
\al{
 \pv\M =  \braket{\WW^\M}{\hh} + \OO(\heta^2), \eqlabel{EOMmu}
}
It turns out that at all times, the drift induced by a small perturbation can be found by taking the overlap integral between the perturbation and the response function $\WW^\M$ corresponding to that degree of freedom. 

This result justifies to call the critical adjoint eigenmodes `response functions' \cite{Biktasheva:2003}. In engineering terms, they are also the spatiotemporal `impulse response' to a localised stimulus. This property can also be used to estimate RFs numerically or in future experiments \cite{Dierckx:2017}. 

In the above-discussed moving frames of reference, the equations of motion become
\alnn{% eq{eomsimple}{}
  \dd_\tp \X^\A &= \vo\A(\mPsi) + \braket{\WW^\A(\mPsi)}{\hh}, \\ 
  \dd_\tp \rPhi &= \romgo(\mPsi) + \braket{\WW^\rPhi(\mPsi)}{\hh}, \\
  \dd_\tp \mPsi &= \mOmgo + \braket{\WW^\mPsi(\mPsi)}{\hh}.
}
 
which can also be summarised as \cite{Dierckx:2017PRL}:
\alnn{
  \dd_\t \X^\M = \vo\M(\mPsi) + \braket{\WW^\M(\mPsi)}{\hh}
}

Note the presence of a zeroth order motion $\vo\A(\mPsi)$, which accounts for spiral tip motion even in the absence of perturbations. Moreover, the rotation rate $\vo\rphi \equiv \romgo$ is allowed to depend on the meander phase $\mPsi$. 

To find the net drift effects, motion generally needs to be time-averaged in the laboratory frame of reference, starting from:
\aleq{EOMlab}{
  \dd_\t \X^\a &= \R\a\A(\rPhi) \vo\A(\mPsi) + \R\a\A(\rPhi) \braket{\WW^\A}{\hh}, \\ 
  \dd_\t \rPhi &= \romgo(\mPsi) + \braket{\WW^\rPhi}{\hh}, \\
  \dd_\t \mPsi &= \mOmgo + \braket{\WW^\mPsi}{\hh}. 
}
 
We will provide some elementary examples in Sec. \ref{sec:applications}. 

At this stage, we can understand why different frames of reference may be used depending on the context. Let us suppose that we describe a process in which the perturbation varies only slightly across the essential support of the RFs of the system, say $\hh(\r) = \E^\a(\r) \HM \dd_\a \uu$,
where $\HM\in\Real^{\Nv\times\Nv}$ is a constant matrix acting in the space of reactive components (state variables).
Then, the overlap integrals will typically be approximated as \cite{Keener:1988, Biktashev:1994, Verschelde:2007, Dierckx:2013}
\al{
   \braket{\WW^\M}{\hh} \approx \E^\a(\Xv) \braket{\WW^\M} {\HM \dd_\a \uu}
}
and for non-homogeneous $\E^\N$ the next-to-leading order will be small only if the response functions $\WW^\M$ are well localised near the point $\Xv$. Although the extent of the RF is fixed by the parameters of the model, the observer can thus describe the reduced system more accurately by an appropriate frame and tip choice. E.g. the interaction of a three-dimensional scroll wave in a detailed cardiac geometry, the results using the tip frame are expected to be more accurate than the center frame description. A similar argument holds for wave fronts, whose critical adjoint eigenmodes are also localised around the wave front.

\section{Geometric interpretation: dynamics on the inertial manifold\label{sec:geom}}

The methods and results presented above can be represented geometrically in the language of dynamical systems on manifolds \cite{Debussche:1991, Biktashev:1996}. 
The formal application of the idealized scheme presented below, as used in this paper as well as, explicitly or implicitly, in some  other previous studies, has well known technical difficulties related to the facts that our phase space is a functional space, \ie\ is infinite-dimensional, and that $\SE2$ is a non-compact group. Some of the implications of these are discussed in the concluding section; for now, we proceed ignoring these technical difficulties, in order to describe interesting physical phenomena, leaving rigorous mathematical treatment for subsequent studies. The first part of this section is an elaboration on the short discussion given in Sec. \ref{sec:symmetries} above. 

\subsection{Phase space orbits}

Every possible state of the system \eq{RDE} at a fixed time, \eg\ $\UU_0(\r)$, can be represented as a point of an infinite-dimensional phase space $\PS$. 
The set $\UU(\r,\t)$ obtained by evolution from the initial condition $\UU(\r,\t_0)=\UU_0(\r)$ according to \eq{RDE} is the \textit{orbit} of $\UU_0$. For the formal development of the theory, we take $\t\in \mathbb{R}$, but we will not use backward time evolution, which will be ill-posed for the systems considered. Of the examples shown in Fig. \ref{fig:patterns}, only rigidly rotating spiral waves and meandering spiral waves with rational $2\pi/\alp$ have closed orbits. 

\subsection{Spatial vs. spatiotemporal symmetries}

Since time is treated as an parameter in phase space dynamics (\ie\ it is not a `direction' of the phase space), it will be useful to distinguish between different symmetry groups that include or not include time symmetry. As in Sec. \ref{sec:symmetries}, we take  $\Group = \SEN \times \mathbb{R}$ and $\Subgroup$ the largest continuous subgroup of $\Group$ which leaves a given orbit invariant. The largest continuous subgroup of $\SEN $ leaving the given orbit invariant is denoted $\SSubgroup$. Then, we have
\al{
 \begin{array}{ccc}
   \SEN & \leq & \Group \\
     \rotleq & & \rotleq \\
   \SSubgroup & \leq & \Subgroup 
\end{array}    
}

The number of broken symmetries $\NBS$, broken space symmetries $\NBSS$ and broken time symmetries $\NBTS$ by the solution are then given by:
\al{
\NBS &=   \dim(\Group) - \dim (\Subgroup), \nn \\
\NBSS &=   \dim(\SEN) - \dim (\SSubgroup), \\ 
\NBTS &=  \NBS - \NBSS. \nn 
}

\subsection{Quotient space}

The \textit{group orbit} of a state $\UU(\r)$ under a group $\AGroup$ is given by
\al{
\AGroup \UU(\r) = \{ \g \UU(\r)\ |\  \g \in \AGroup \} .
}
It will be useful to study the dynamics of the system, disregarding spatial Euclidean symmetries. Therefore, we consider equivalence classes: $\UU_1 \sim \UU_2$ iff there exists $\g \in \SEN$ such that $\UU_1 = \g \UU_2$. The space of equivalence classes is called the quotient space, denoted $\Quot = \PS / \SEN $ and contains as members the group orbits under $\SEN$.

 The original dynamical system \eq{RDE} induces a (reduced) dynamics on the states in $\Quot$. 

The equilibrium points of the dynamics in $\Quot$ are termed \textit{relative equilibria} in $\PS$, as they correspond to equivalence classes under the action of the Galilean group in $\PS$. Hence, from the examples discussed above, the rigidly translating waves and rigidly rotating spirals are relative equilibria of the reaction-diffusion system. (For comparison, the resting state is a `true' equilibrium). 

The limit cycles of the dynamics in $\Quot$ are termed \textit{relative periodic orbits} in $\PS$. Of the examples above,  modulated traveling waves in one spatial dimension, biperiodically meandering spiral (both flower-like and star-like) and spirals in resonant meander are relative periodic orbits. 

We will below further describe solution to the RD system
 that are relative equilibria or relative periodic orbits. These have an attractor $\Att \subset \Quot$ of dimension 0 or 1. The attractor is not necessarily unique to the medium. For instance, for the case of a 2D medium supporting spiral waves, one already finds 4 different attractors: the resting state, a traveling plane wave and spiral waves of two opposite chiralities.

\subsection{Inertial manifold}

The presence of stable persisting patterns in the system puts a special structure on the phase space. Consider the set of states in $\PS$ that belong to the same relative equilibrium or relative periodic orbit:
\al{
  \MM = \Group \Att = \{ \g \UU(\r)\ |  \UU(\r) \in \Att, \,  \g \in \Group \} .
}
This set $\MM$ (implicitly depending on a given choice for $\Att$) forms a manifold in the phase space $\PS$. 
From our previous assumption that $\linop$ has no eigenvalues with positive real part, it follows that nearby trajectories are attracted with exponential speed to $\MM$, whence it is an \textit{inertial manifold} \cite{temam:1988}~\footnote{
  Technically speaking, this is not precisely true: there are continuous branches of the spectrum that reach the imaginary axis, as a result of which the attraction is not exponential. However, on the ``physical level of rigour'', these continuous branches correspond to large-scale perturbations of the periphery of the spiral wave, so if we consider a finite vicinity of the spiral core, the approach is still sound. This is one of the outstanding technical difficulties we mentioned earlier. 
}. Since unidirectional attraction is implied here, the emergent structure of an inertial manifold is typical for dissipative systems.  

%In our case, the inertial manifold emerges as the image of a prototypical solution $\UU(\r,\t)$, such as a traveling wave or a spiral wave, under the spatial action of the orientation-preserving isometries:
%\al{
% \MM = \{ \Group \UU(\r,\t)\ |\  \Group \in \SE\Ns, \; \t\in\Real \} .
%}
A given system \eq{RDE} may have various inertial manifolds corresponding to different prototypical solutions $\UU$, such as the resting state, a propagating pulse, or a spiral wave with either chirality. 

Inertial manifolds constitute a special case of slow manifolds \cite{Debussche:1991}. When the Euclidean symmetry is not broken ($\hh = \bzero$), the dynamics on the inertial manifold is trivial and given by the orbit of that solution. In the presence of a perturbation, however, the solution will not follow the original orbits anymore, but slowly drift from one original orbit to the other. This regime has been the starting point for many analytical studies of non-linear wave dynamics, \eg\  \cite{Kuramoto:1980, Keener:1986, Henry:2004, Dierckx:2011, Dierckx:2013, Lober:2014}, including extension to 3D scroll waves \cite{Keener:1988, Biktashev:1994, Henry:2002, Verschelde:2007, Biktashev:2010, Dierckx:2017PRL}. 

A remarkable property is that in the high (infinite)-dimensional phase space, the inertial manifold associated to the patterns above is finite-dimensional. Given that the orbit $\UU(\r,\t)$ has dimension $1$ and $\dim(\SE\Ns) = \Ns(\Ns+1)/2$, the maximal dimension of $\MM$ would be $\Ns(\Ns+1)/2 + 1$. However, a state $\UU(\r,t_0)$ may be invariant under a subgroup $\J$ of $\SE\Ns$, in which case $\J$ is called an \textit{isotropy subgroup}.  E.g. a plane wave in 2D is invariant under translations along its wavefront. 

The elements of $\J$ correspond to the preserved symmetries of the solution, such that the number of broken spatial symmetries amounts to:
\al{
  \NBSS = \dim(\SE\Ns) - \dim(\J).
}

For solutions that are relative equilibria, time evolution can be represented by a Euclidean group action: $\UU(\r,t) = \g \UU(\r,0)$ with $\g=\g(\t)\in \Group$ for all $\t\in\Real$. In this case, the time symmetry is not explicitly broken, and we say that $\NBTS=0$. For relative periodic orbits, $\UU(\r,\t) = \g(\t) \UU(\r,0)$ with $\g(\t) \in \Group$ only when $\t$ is an integer multiple of the period $\To$ of the limit cycle in $\Quot$. In this case, we say that the number of broken time symmetries $\NBTS = 1$. 

From the above, it follows that the dimension of the inertial manifold equals the number of broken continuous symmetries, \ie\ 
\al{
  \dim (\MM) = \NBS =  \NBSS + \NBTS. 
}

\subsection{Collective coordinates parameterize the inertial manifold}

%The inertial manifold of dimension $\NBS$ can be parameterized by a set of $\NBS$ labels, known as collective coordinates; we denoted these above as $\X^\M$. Since the inertial manifold acts as a local attractor in phase space, the dynamics within the vicinity of the inertial manifold can be approximated by a system of $\NBS$ coupled ordinary differential equations. 
%\vnbcmt{variant:}

By definition, the inertial manifold is flow-invariant, so the dynamics on it can be described by a system of $\NBS$ coupled ordinary differential equations. Since it is a local attractor, this system will approximate the dynamics within the vicinity of that manifold.

The different frames shown in this work serve the same purpose: to uniquely characterize a dynamical state of the system. Different frame choices thus correspond to different parameterizations of the $\NBS$-dimensional inertial manifold. 

Since every spatial collective coordinate is related to a broken symmetry, one has that
\al{
 \df{\UU}{\X^\M} = - \df{\UU}{\xp^\M},
}
where $\xp^\M$ is a co-moving frame coordinate and the prototypical solution $\UU(\x,\t)$ is assumed to depend on the collective coordinates $\X^\M$ as parameters. 
E.g. since a traveling wave has the shape 
$\UU(\x,\t) = \uuo(\x - \X(\t))$, one has 
\al{
\df{\UU}{\X} = - \df{\uuo}{\x} . 
}
Similarly, for spiral waves 
\al{
\df{\UU}{\X^\A} = - \df{\uuo}{ \xp^\A}, \qquad \df{\UU}{\rphi} = - \df{\uuo}{ \polang},
}
where $\polang$ is the polar angle in the chosen frame of reference and $\rphi$ is the spiral's rotation phase. 

\subsection{Tangent spaces}

One can introduce tangent vectors to the inertial manifold by differentiating the state with respect to a collective coordinate:
\al{
  \VV_\M = \df{}{\X^\M}  \UU\left(\g^{-1}(\X)\r,\t\right). 
} 
Here, $\g^{-1}(\X)$ has the effect of working in the quotient space (or fully co-moving frame). 

If one differentiates with respect to the time-like collective coordinate ($\mPsi$), one obtains a tangent vector to the orbit of the state under time evolution according to the RDE \eq{RDE}. 

These tangent vectors to the inertial manifold thus correspond to the Goldstone modes (GM) of the theory, or right-hand critical modes of the system, also called $\VV_\M$ in sections \ref{sec:linear}-\ref{sec:drift}. The non-critical modes of the system then point towards the other directions in phase-space, off the inertial manifold. 

In every point of the inertial manifold, one can also introduce a set of dual vectors $\WW^\M$, that correspond to the response functions defined and used above. In every point of the inertial manifold it is possible to choose
\alnn{ % \eqlabel{WW-VV-short-biort}
 \braket{\WW^\M}{\VV_\N} = \kron\M\N.
}

\subsection{Effect of a small perturbation}

Now, suppose one starts with a pattern state that exactly equals the prototypical solution $\UU$ 
to the RDE.  Hence the initial dynamics of this state is on the inertial manifold. Then, applying a small stimulus $\hh$ at time $\t=0$ has the effect of shifting the state in phase space.
Unless $\hh$ is carefully tuned, 
the state after applying the perturbation does not lie on the inertial manifold anymore. However, we assume that the wave solution $\UU$ is dynamically stable, such that  the resulting state evolves towards the inertial manifold and converges to an orbit in it. 

Then, the resulting orbit will generally not be the same state as would have been reached without the perturbation. That is, one retrieves the same solution (spiral or traveling wave), but with different collective coordinates $\X^\M$. A practical way of finding this shift, is to measure which part of the perturbation was tangential to the inertial manifold, and that is precisely what Eqs. \eq{EOMmu} mean. The part of the perturbation that is orthogonal to the inertial manifold will decay over time, and only represent transient dynamics (as long as the amplitude of the perturbation is small enough). 

\subsection{Coordinate change in the tangent manifold \label{sec:coordTF} }

Geometrically speaking, the different coordinate systems which we presented in Sections \ref{sec:1D} and \ref{sec:2D} are alternative ways to define collective coordinates, and therefore alternative bases vectors for the tangent manifold. With that respect, we can state that 
\begin{align}
 \cen{\VV}_\M &= \frac{\dd \cen{X}^\N}{\dd \tip{X}^\M} \tip{\VV}_\N, \\
 \cen{\WW}^\M &= \frac{\dd \cen{X}^\M}{\dd \tip{X}^\N} \tip{\WW}^\N. 
\end{align}
such that 
\begin{align}
   \cen{\VV} &= \Jac \tip{\VV}, \\ 
   \tip{\WW} &= \JacH \cen{\WW}, \eqlabel{adjoint}
\end{align}
where $\Jac$ is the Jacobian of the coordinate transformation and $\JacH$ is its Hermitian transpose. Eq. \eq{adjoint} is an example of the adjoint representation, and will be exemplified by Eqs. \eq{tfV}, \eq{tfW}. 

\section{Application: spiral drift in a constant external field \label{sec:applications}}

\subsection{Motivation}
A simple perturbation field that can be considered is a constant vector field that couples to the gradient in one or more state variables: 
\al{
   \hh = \vE \cdot \vec{\nabla} \Mb \uu. \eqlabel{EFD}
}
In chemical systems, this term can model an applied electrical field $\vE$ that induces drift of different ions $u_j$ with respective mobilities $\Mel{}\j$ that are found on the diagonal of the matrix $\Mb$. The resulting phenomenon in a chemical context is known as `electroforetic drift'. 

In 3D systems supporting spiral waves, the motion induced by a small filament curvature $\kur$ also generates a term of the form \eq{EFD}, with $\vE = \kur\vN$ and $\Mb = \HP$, where $\vN$ is the unit normal vector to the filament curve. 

\subsection{Average drift of a circular core spiral: center versus tip frame}

For circular core spirals, the law of motion \eq{EOMlab} with the stimulus \eq{EFD} yields
\aleq{EFDcirc}{
 \dd_\t \X^\a &= \R\a\A(\rPhi) \vo\A + \R\a\A(\rPhi) \Mel\A\B \R\B\b(\rPhi) \E^\b \\
\dd_\t \rPhi &= \romgo + \Mel\rPhi\A \R\A\a(\rPhi) \E^\a
}
 
with 
\al{
  \Mel\M\B &= \braket{\WW^\M }{\Mb \VV_\B}. \eqlabel{MABcirc}
}

The relations \eq{EFDcirc} hold both in the tip frame (with $\vo\A$ constant) and the center frame (with $\vo\A=0$). However, the RFs and Goldstone modes are different in both cases, such that in the end both descriptions are equivalent, as we now show. 

Recasting the second equation for $\t(\rPhi)$ instead of $\rPhi(\t)$, we immediately get
\al{
\t(2\pi) - \t(0) = \To + \OO(\E^2)
}
since the average of a rotation matrix over its angle vanishes. Next, dividing the first equation of \eq{EFDcirc} by the second, we obtain
\al{
  \romgo \dd_\rPhi \X^\a =&  \R\a\A(\rPhi) \vo\A +  \mbox{} \\
  & \R\a\A(\rPhi) \left( \Mel\A\B  - \frac{\vo\A \Mel\rPhi\B}{\romgo}\right) \R\B\b(\rPhi) \E^\b 
  + \OO(\E^2) . \nn 
}
Integrating over $\rPhi$ on $[0, 2\pi]$ delivers the period-averaged drift velocity:
\al{
  \V^\a & = \frac{\X^\a(2\pi) - \X^\a(0)}{\t(2\pi) - \t(0)} \nn \\
  &=  \left( \Mel\A\B  - \frac{\vo\A \Mel\rPhi\B}{\romgo}\right) 
  \frac{\left(\kron\a\b \kron\B\A + \lcsf\a\b \lcsf\B\A \right)}{2} \E^\b \nn \\ 
  &=\gone \E^\a + \gtwo \lcsf\a\b \E^\b \eqlabel{EOM_EFD}
}
with 
\al{ 
\gone &= \frac{1}{2}\left( \Mel\A\A  - \frac{\vo\A \Mel\rPhi\A }{\romgo}\right),  \nn \\
\gtwo &= \frac{1}{2}\lcsf \B\A \left( \Mel\A\B  - \frac{\vo\A \Mel\rPhi\B}{\romgo}\right). 
}
In the center frame, with $\vo\A=0$, this is the classical result from \cite{Biktashev:1994}. However, does the calculation in the tip frame yield the same result?

Here, we can make the discussion of Par. \ref{sec:coordTF} explicit. The center and tip frames are for a circular-core spiral  related by
$  \tip\xp^\A = \cen\xp^\A - \dfil^\A. $
 Call the polar angle in either frame $\cen{\polang}$ and $\tip{\polang}$; the unperturbed spiral solution is $\cen\uuo(\cen\xp^\A) = \tip\uuo(\tip\xp^\A)$. Then, since $\dd_{\tip{\polang}} = -\lcsf\B\A \tip\xp^\A \dd_\B \tip\uuo,
$ it follows that 
% \al{ \eqlabel{tfV}
% \left(
% \begin{array}{c} \tip{\VV}_\A(\tip{\rho}^\C) \\ \tip{\VV}_\rphi(\tip{\rho}^\C) \end{array}\right) = \left(
% \begin{array}{cc} \mathbf{1} &\mathbf{0} \\ \eps^\A_{\hs \B} \dfil^\B & 1 \\ \end{array}\right) \left(
% \begin{array}{c} \cen{\VV}_\A(\cen{\rho}^\C) \\ \cen{\VV}_\rphi(\cen{\rho}^\C) \end{array}\right). 
% }
\al{ \eqlabel{tfV}
  \Mx{
    \tip{\VV}_\A(\tip{\rho}^\C) \\ 
    \tip{\VV}_\rPhi(\tip{\rho}^\C) 
  }
  =
  \Mx{
    \bone & \bzero \\
    \lcsf\A\B \dfil^\B & 1
  } \Mx{
    \cen\VV_\A\left(\cen\xp^\C\right) \\ 
    \cen\VV_\rPhi\left(\cen{\rho}^\C\right)
  }. 
}

To find the RF transformation law, we consider a localized perturbation of the $\j$-th state variable at a given position: 
%$\h_\j = \heta \dirac(\tip\xp^\A - \tip\xp^\A_0) = \heta \dirac(\cen\xp^\A - \cen\xp^\A_0)$. 
$\hh=\left(\h^\k\right)$, 
$\h^\k = \heta \dirac(\xp^\A - \xp^\A_0) \dirac(\t) \kron\k\j$.
From \eq{EOMlab}, we find in the center frame
\al{
  \dd_\t \cen\X^\a &= \heta \R\a\A(\cen\rphi) \dirac(\t) \cen\W^\A_\j(\cen\xp^\A),  \eqlabel{EFDcirc2c} \\
  \dd_\t \cen\rphi &= \romgo + \heta \cen\W^\rPhi_\j(\cen\xp^\A) \dirac(\t), \nn 
}
and in the tip frame
\al{
  \dd_\t \tip\X^\a &= \R\a\A(\tip\rphi) \vo\A +  \heta \R\a\A(\tip\rphi) \tip\W^\A_\j(\tip\xp^\A) \dirac(\t),  \eqlabel{EFDcirc2t} \\
  \dd_\t \tip\rPhi &= \romgo + \heta \tip\W^\rphi_\j(\tip\xp^\A) \dirac(\t). \nn 
} 

Differentiating $\tip\X^\a = \cen\X^\a + \R\a\A(\cen\rphi)\dfil^\A $ with respect to time delivers
\al{
  \dd_\t \tip\X^\a &= \dd_\t \cen\X^\a - \R\a\A(\cen\rphi) \lcsf\A\B \, \dd_\t \cen\rphi \, \dfil^\B,  \eqlabel{Xtipcenter} \\
  \dd_\t \tip\rphi &= \dd_\t \cen\rphi . \nn 
}
After substituting Eqs. \eq{EFDcirc2c} and \eq{EFDcirc2t} into Eq. \eq{Xtipcenter}, we conclude from the zeroth order term in $\eta$ that
\al{
  - \lcsf\A\B \dfil^\B \romgo = \vo\A \eqlabel{defVA}
}
and from the first order term in $\eta$ that
% \al{ \eqlabel{tfW}
% \left(
% \begin{array}{c} \tip{\WW}^\A(\tip{\rho}^\C) \\ \tip{\WW}^\rPhi(\tip{\rho}^\C) \end{array}\right) = \left(
% \begin{array}{cc} \bone & - \lcsf \A \B \dfil^\B \\ \bzero  & 1  \end{array}\right) \left(
% \begin{array}{c} \cen{\WW}^\A(\cen{\rho}^\C) \\ \cen{\WW}^\rPhi(\cen{\rho}^\C) \end{array}\right). 
% }
\al{ \eqlabel{tfW}
  \Mx{ \tip{\WW}^\A(\tip{\rho}^\C) \\ \tip{\WW}^\rPhi(\tip{\rho}^\C) }
  = 
  \Mx{ \bone & - \lcsf \A \B \dfil^\B \\ \bzero  & 1 }
  \Mx{ \cen{\WW}^\A(\cen{\rho}^\C) \\ \cen{\WW}^\rPhi(\cen{\rho}^\C) } .
}
 
Hence, using Eq. \eq{defVA} we can verify that
\al{
\tip{\gone} &= \braket{\tip\WW^\A - \frac{\vo\A}{\romgo} \tip\WW^\rPhi}{\Mb \tip\VV_\A} \eqlabel{gonedouble}\\
&= \braket{\cen\WW^\A - \lcsf \A \C \dfil^\C \cen\WW^\rPhi - \frac{\vo\A}{\romgo} \cen\WW^\rPhi}{\Mb \cen\VV_\A} = \cen{\gone} \nn 
}
and similar for $\gtwo$. This result resolves an apparent paradox: the expression for the filament tension $\gone$ is different in the center frame and the tip frame, as shown by the first and second line of Eq. \eq{gonedouble}. However, both expressions are equal since the RFs and GMs appearing in these expression differ for both frames.  The final result is thus independent of the chosen frame of reference. 

\subsection{Phase-locking of a meandering spiral under a constant external field. }

%\begin{figure} \centering
%\raisebox{2cm}{a)} \includegraphics[width=0.25\textwidth]{example.jpg}\\
%\caption{Figure with phase-locked trajectories}
%\end{figure}

In systems where spiral waves are found where the angle $\alp$ between subsequent petals of the meander flower is close to a simple fraction of $2\pi$, the rotation phase may lock to the external field if that is strong enough~\cite{Dierckx:2017PRL}.

To capture this phenomenon quantitatively, we take the minimally rotating finite-core frame of Fig. \ref{fig:frames_meander}b. Then, when $\E= \abs{\vE} = 0$, the $\X^\b(\t)$ describes the motion of a point moving on a circle with radius equal to the mean radius of the meander flower. To recover the precise tip motion from $\X^\b(\t)$ one can take
\al{
 \ptip^\b(\t) = \X^\b(\t) + \R\b\B (\rPhi(\t)) \dfil^\B(\mPsi(\t)). 
}  
Plugging Eq. \eq{EFD} into Eq. \eq{EOMlab} then delivers: 
\aleq{EOMEFD}{
  \dd_\t \X^\a &= \R\a\A(\rPhi) \vo\A + \R\a\A(\rPhi) \Mel\A\B(\mPsi) \R\B\b(\rPhi) \E^\b, \\ 
  \dd_\t \rPhi &= \romgo + \Mel\rPhi\B(\mPsi) \R\B\b (\rPhi) \E^\b, \\
  \dd_\t \mPsi &= \mOmgo + \Mel\mPsi\B(\mPsi) \R\B\b (\rPhi) \E^\b.  
}
 
The matrix element $\Mel\M\B$
is formally defined as in Eq. \eq{MABcirc}, but now $\M \in \{\x,\y,\rPhi, \mPsi \}$. Note that 
functions
$\Mel\M\B(\mPsi)$ and $\R\a\A(\rPhi)$ are all $2\pi$-periodic. Due to our choice of coordinate system, $\vo\A$ and $\romgo$ are non-zero but constant. 

In our analysis, we first treat the case where $\alp \ll 2\pi$, \ie\ near the $1:1$ resonance. Then, we assume that both $\romgo$ and $\E = \abs{\vE}$ are $\OO(\heta)$, such that a locked solution may exist. Under those conditions, $\mPsi(\t)$ will remain a monotonous function, and we can trade the time coordinate $\t$ for $\mPsi$. Expanding up to linear order in $\heta$ delivers:
\algrp{PL}{
  \mOmgo \dd_\mPsi \X^\a &= \R\a\A(\rPhi) \vo\A + \OO(\heta),                                   \eqlabel{PLX} \\ 
  \mOmgo \dd_\mPsi \rPhi &= \romgo + \Mel\rPhi\B(\mPsi) \R\B\b(\rPhi) \E^\b + \OO(\heta^2).     \eqlabel{PLphi}
}

If there is phase-locking of the rotation angle $\rPhi$ to the meandering phase $\mPsi$, then equation \eq{PLphi} will give a solution for $\rPhi$ that remains in a bounded vicinity of some mean value ${\base\rPhi}$,
\alnn{
  \rPhi(\mPsi) = {\base\rPhi} + {\pert\rPhi}(\mPsi),
  \qquad
  \abs{{\pert\rPhi}(\mPsi)}<\const.
}
We will be looking for the simplest case of 1:1 phase locking, such that
\alnn{
  \rPhi(\mPsi+2\pi) \equiv \rPhi(\mPsi).
}

From Eq. \eq{PLphi}, the rate-of-change of $\rPhi$ is also small (\ie\ $\OO(\heta)$), while the rate-of-change of $\mPsi$ is $\mOmgo = \OO(1)$. Therefore, we can assume that ${\pert\rPhi} =\OO(\heta)$ during one meander cycle, and average over the cycle. Denoting $\rPhi(2\pi\nm)$ as $\rPhi_\nm$, we obtain the difference equation:
\al{
  \frac{\rPhi_{\nm+1} - \rPhi_\nm}{\To} = \romgo + \avg{\Mel\rPhi\B} \,  \R\B\b(\rPhi_\nm) \E^\b. \eqlabel{condition}
}
Then, a phase-locked angle is a fixpoint of Eq. \eq{condition}. Without loss of generality, we can take $\vE = \E \ex$ here, to show that this condition can be formulated as:
\al{
  \romgo + \Aconst\E\cos({\base\rPhi} - \aconst) = 0 \eqlabel{coseq}
}
with
\alnn{
   \Aconst \e^{-\iu\aconst} = \avg{\Mel\rPhi1} + \iu \avg{\Mel\rPhi2} .
}
 
A necessary condition for a phase-locked solution to exist is then $|\E\Aconst/\romgo| >1$, \ie\ ~{\cite{Dierckx:2017PRL}}
\alnn{
  \abs{\E} > \Ecrit = \frac{\romgo}{\Aconst} + \OO(\heta^2). % \eqlabel{ecrit} 
}

From Eq. \eq{coseq}, the rotation angle around which the locked equation will equilibrate is
\alnn{
 {\base\rPhi} = \aconst + \arccos\left( \frac{\romgo}{\Aconst\E}\right). 
}

The second solution to Eq. \eq{coseq}, namely $\aconst - \arccos\left( \romgo / {\Aconst\E} \right)$, is unstable. 

Since in the phase-locked state, $\rPhi = {\base\rPhi} + \OO(\heta)$, Eq. \eq{PLX} can be readily integrated to find the net spatial displacement during a phase-locked meander cycle: 
\al{
  \X^\a(2\pi) - \X^\a(0) = \R\a\A({\base\rPhi})\vo\A \To  + \OO(\heta),
}
where $\To=2\pi/\mOmgo$. 
From $\dd_\t \mPsi = \mOmgo + \OO(\heta)$, one finds moreover that the time needed to make a full meander cycle is $\T + \OO(\heta)$, such that the net drift speed during phase locking is
\aleq{PLVnet}{
   \V^\a &= \frac{\X^\a(2\pi) - \X^\a(0)}{\To} + \OO(\heta) \\
         &= \R\a\A({\base\rPhi}) \vo\A + \OO(\heta). 
}

\begin{figure}\centering
 \includegraphics[width = 0.45 \textwidth]{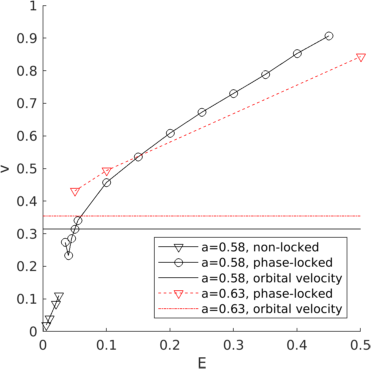}
\caption[]{\label{fig:PLorder1} Magnitude of spiral wave drift velocity in Barkley's model ($\bpar=0.05$, $\cpar = 0.02$), far from resonance ($\a=0.58$, black), and close to resonance ($\apar=0.63$, red). Close to the phase-locking threshold, the drift speed approaches the orbital velocity $\c=\romgo\rt$, with $\rt$} the average radius of the meander flower. 
\end{figure}

In words, it can be concluded that the drift speed of a meandering spiral phase-locked to a constant vector field $\vE$ is of the order of the `orbital' velocity $\vo{}$, \ie\ the net speed at which the tip traverses the rim of the meander flower. Importantly, within the locking region \emph{the drift velocity is not proportional to the perturbation strength $\E$!}
In terms of the velocity parallel or perpendicular to the applied field, Eq. \eq{PLVnet} reads:
\alnn{
   \V^{\parallel} &= \vo{} \cos \base\rPhi + \OO(\heta), &
   \V^{\perp} &= - \vo{} \sin \base\rPhi + \OO(\heta),
}
a result which was already quoted in \cite{Dierckx:2017PRL}.  The magnitude of drift velocity is therefore expected to be $\V = \c + \OO(\heta)$. Fig. \ref{fig:PLorder1} confirms that this is the case for simulations in Barkley's model. 

Until here, we have described phase-locking of a single meander cycle to an external field, which happens when $\abs{\romgo} \ll \mOmgo$. Since $\mOmgo = 2\pi/\To$ and $\romgo = \alp/\To$ with $\alp$ the angle between subsequently visited petal tips, phase-locking is found generally when 
\alnn{
 \alp \approx 2\pi\Mint , \qquad \Mint\in\Zahlen.
}
In some cases, a higher-order resonance may occur, \ie
\alnn{
  \alp \approx 2\pi\Mint/\Nint , \qquad \Mint,\Nint\in\Zahlen, \; \Nint\ne0
}
(naturally, practical interest present cases of small $\Mint,\Nint$). 
In this case, $\Nint$ meander cycles can be re-interpreted as a single cycle to which the theory \eq{EOMlab}-\eq{PLVnet} can be applied, with $\romgo$ replaced by $\Nint\romgo-\Mint\mOmgo$. 

\subsection{Regime close to phase-locking}

It could happen that the applied field 
$\vE$ is not strong enough to enforce phase-locking, which occurs when $\abs{\E}<\Ecrit$.
In that case, the net drift can be found from  the instant equations of motion \eq{EOMlab} by averaging of one sort 
or another. In the case of circular core spirals, this was done by sliding time averaging~\cite{Biktashev:1995b} or by imposing condition of periodicity on the perturbed solution that in turn requiring a solvability condition~\cite{Biktasheva-etal-2010}. Since for meandering spirals the unperturbed solution is not periodic but biperiodic, the situation here requires a bit more care. In~\cite{Dierckx:2017} we considered the case below but not too close to phase-locking threshold, and used Fourier series. For the case $\abs{E}\lesssim\Ecrit$, this does not work, so we resort to the more robust method, by sliding averages. First we do averaging over the meander phase:
\alnn{
  \slavg{\rPhi}(\mPsi) = \frac{1}{2\pi}\int_{\mPsi-\pi}^{\mPsi+\pi} \rPhi(\mPsi') \,\d\mPsi'.
}

Then, if $|\romgo| \ll \mOmgo$, 
\al{
 \dd_\t \slavg{\rPhi} = \avg{\romgo} + \avg{\Mel\rPhi\A} \R\A\a(\slavg{\rPhi}) \E^\a + \OO(\E^2). 
 \eqlabel{bigPhi}
}
As above, we can write the second term in the right-hand side as $\E\Aconst\cos(\rPhia-\aconst)$. 
Let us denote the time needed to complete a full meander flower as
\alnn{
 \Troto = \frac{2\pi}{\avg{\romgo}} = \frac{2\pi}{\alp} \To.
}
 
Under a perturbation this value changes to $\Trot$. Then, straightforward integration of Eq. \eq{bigPhi} delivers:
\alnn{
 \Trot = \frac{\Troto}{\sqrt{1 - \left(\E/\Ecrit\right)^2}}. 
}
 
Therefore, if $\E$ approaches $\Ecrit$ from below, the time needed to complete a full meander flower diverges. Beyond $\Ecrit$ the spiral fails to complete a meander cycle and is phase-locked.  

\subsection{Average drift speed of a non-phase-locked meandering spiral}

\begin{figure} \centering
\includegraphics[width=0.5                                                                                                     \textwidth]{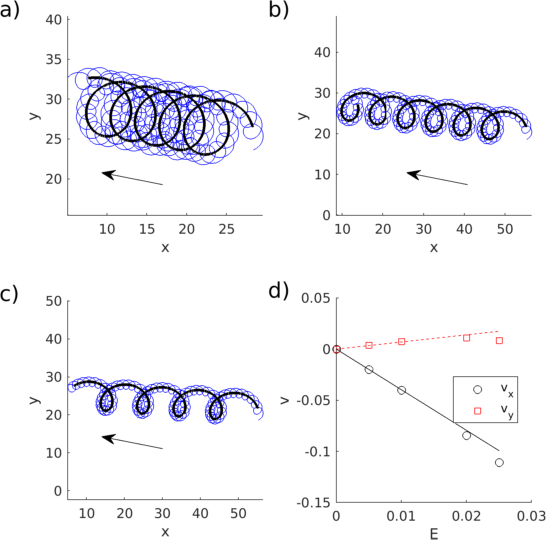}\\
\caption[]{Non phase-locked drift of a meandering spiral in Barkley's model due to a constant external field $\vE = \E\ex$, for different field strengths: (a) $\E = 0.01$, (b) $\E=0.02$, (c) $\E=0.025$. Model parameters as in Fig. \ref{fig:patterns}d and $\HM = \HP$. Panel d) shows the parallel and perpendicular components of drift (data points), which closely match the theoretical predictions given by Eq. \eq{Vmean} (lines). \label{fig:freeEFDBa} }
\end{figure}

\begin{figure} \centering
\includegraphics[width=0.5 \textwidth]{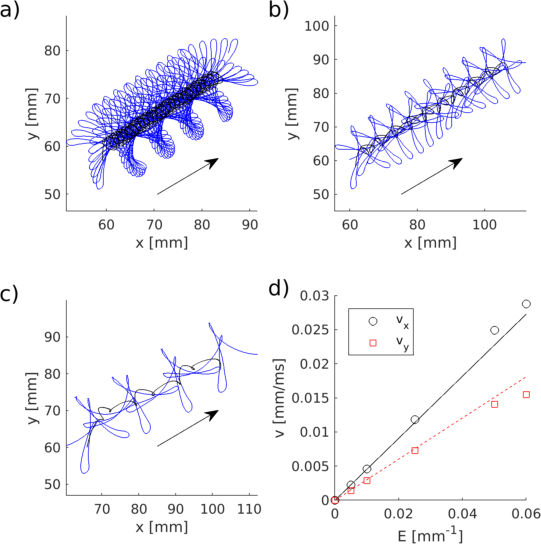}\\
\caption[]{Same as Fig. \ref{fig:freeEFDBa}, but for the Fenton-Karma guinea pig cardiac model from Fig. \ref{fig:patterns}, which has a linear core. Applied field strengths are (a) $\E= 0.005\,\millim^{-1}$, (b) $\E= 0.025\,\millim^{-1}$, (c) $\E=0.06\,\millim^{-1}$.
\label{fig:freeEFDFK}
}
\end{figure}

We continue our analysis of Eq. \eq{EOMlab} in the regime where phase-locking does not occur, \ie\ when $\alp / 2\pi$ is not a simple fraction or when $E$ is sufficiently small, say $\E = \OO(\heta)$.  

For any of the reference frames shown in Fig. \ref{fig:frames_meander}, we can  in a first step integrate Eqs. \eq{EOMEFD} to find the net displacement in space, time and rotation angle during the single meander cycle, say the $\nm$-th. At the start of this cycle, taking place at $\t=\t_\nm$ the collective coordinates of the spiral are denoted $\X^{\a}_{\nm}, \rPhi_\nm, \mPsi_\nm$. 

%Thereafter, an iterated map can be used to find the displacement (drift) over many meander cycles. 
%We first define the net displacements in the unperturbed case, with label `u':

For $\t \in [\t_n, \t_{n+1}]$, we can write:
\alnn{
 \mPsi(\t) &= \mPsi_\nm + \base\mPsi(\t) + \pert\mPsi(\t)\\
 \rPhi(\t) &= \rPhi_\nm + \base\rPhi(\t) + \pert\rPhi(\t) \\ 
{ \X^\a(\t)} &=  \X^{\a}_{\nm} + \base\X^\a(\t) +\pert\X^\a(\t). 
}
 
In these right-hand sides, the first term is the initial value at the start of the $\nm$-th cycle, the second term is the evolution in the unperturbed case, and the third term the correction to it, of $\OO(\E)$. Without loss of generality, we can set the $\nm$-th cycle to start at $\mPsi = \mPsi_n = 2\pi\nm$.

The unperturbed evolution is given by
\alnn{
  \base\mPsi(\t) &= \mOmgo (\t-\t_\nm) \nn \\ 
  \base\rPhi(\t) &= \int_{\t_\nm}^\t \romgo({\base\mPsi(\t')}) \,{\d\t'}  \\
 {\base\X^\a}(\t) &= \int_{\t_\nm}^\t  \R\a\A({\base\rPhi(\t')}) \vo\A({\base\mPsi(\t')}) \,{\d\t'} 
}
Due to the group structure of rotations, we can write
\aleq{rot3}{
  \R\A\a(\rPhi) &= \R\A\alpx({\base\rPhi})\R\alpx\betx({\pert\rPhi})\R\betx\a(\rPhi_n), \\
  \R\alpx\betx({\pert\rPhi}) &= \kron\alpx\betx\cos {\pert\rPhi} + \lcsf\alpx\betx \sin {\pert\rPhi}
  = \kron\alpx\betx + \lcsf\alpx\betx  {\pert\rPhi} + \OO(\heta^2). 
}
 
We will below use
\aleq{Malphabeta}{
 {\Mel\mux\alpx(\base\mPsi(\t))} &= {\Mel\mux\A(\base\mPsi(\t)) \R\A\alpx(\base\rPhi(\t)), \qquad \mux \in \{\rPhi, \mPsi\} } \\ 
  \Mel\alpx\betx({\base\mPsi}(\t)) &= \R\alpx\A({\base\rPhi}(\t)) \Mel\A\B({\base\mPsi}(\t)) \R\A\alpx({\base\rPhi}(\t)), \\
  \vo\alpx({\base\mPsi}(\t)) &= \R\alpx\A({\base\rPhi}(\t)) \vo\A(\mPsi)
}
 
Note that these functions are not periodic in $\mPsi$ anymore.

With Eqs. \eq{rot3}-\eq{Malphabeta}, the evolution equation for $\mPsi$ becomes 
\al{
{\dd_\t} \mPsi = \mOmgo + \Mel\mPsi\betx({\base\mPsi}(\t)) \R\betx\b(\rPhi_\nm) \E^\b + \OO(\heta^2) . \eqlabel{dtpsicycle}
}

Since $\rPhi_\nm$ is constant, Eq. \eq{dtpsicycle} yields the duration of the meander period in the presence of a constant external field $\vE$:
\aleq{Tn}{
  \To_\nm &= \t_{\nm+1} - \t_\nm =  \int_{\t_\nm}^{\t_{\nm+1}}  \,\d\t 
  = \int_0^{2\pi} \frac{\d\mPsi}{\mOmgo + \Mel\mPsi\alpx \R\alpx\a(\rPhi_\nm) \E^\a} \\  
  &= \To \left(1 - \mOmgo^{-1} \avg{\Mel\mPsi\alpx} \R\alpx\a (\rPhi_\nm) \E^\a\right)+ \OO(\heta^2).
}
 
%This relation describes how an external field changes the duration of the meander period. 
Note that the resulting period depends on $\rPhi_\nm$, \ie\ on the orientation of the trajectory relative to the applied field $\vE$.

The rotation phase during one cycle can be found by integrating $\dd_\mPsi \rPhi = \dd_\t \rPhi / \dd_\t \mPsi$:
\alnn{%eq{alphan}{}
  \rPhi(\mPsi) - \rPhi_\nm  
  &= \int_{\mPsi_\nm}^{\mPsi} \frac{\romgo(\mPsi) + \Mel\rPhi\alpx(\mPsi) \R\alpx\a(\rPhi_\nm) {\E^\a}}{\mOmgo + \Mel\mPsi\alpx(\mPsi) \R\alpx\a(\rPhi_\nm) {\E^\a}}\d\mPsi \\
  &= {\base\rPhi} +  \NN{\rPhi}{\alpx} \R\alpx\a(\rPhi_\nm) {\E^\a} + \OO(\heta^2),
}

where
\alnn{
  \NN{\rPhi}{\alpx}(\mPsi) &= \mOmgo^{-1} \int_{\mPsi_\nm}^\mPsi \dalp\alpx(\mPsi')  \,\d\mPsi' , \\
  \dalp\alpx(\mPsi) &= \Mel\rPhi\alpx (\mPsi)  -\mOmgo^{-1} \To\romgo \,  \Mel\mPsi\alpx (\mPsi).
}
Evaluation at $\mPsi_{\nm+1}$ delivers the change in rotation angle over one meander period: 
\aleq{alphan}{
  \alp_\nm &= \rPhi_{\nm+1} - \rPhi _\nm \\
  &= \alp +  \To \, \avg{\dalp\alpx} \, \R\alpx\a(\rPhi_\nm) \E^\a + \OO(\heta^2),
}
 
Hence, an applied field will change the angle between consecutively visited petals of the meander flower. 

Finally, we can find the net spatial displacement during a meander cycle:
\alnn{
 \dist^{\a}_{\nm} =&  \X^{\a}_{\nm+1} - \X^{\a}_{\nm} \\
  =&\dist^\a + \To  \R\a\alpx(\rPhi_\nm) \avg{\mel\alpx\alp} \R\alp\b (\rPhi_\nm) \E^\b + \OO(\heta^2),
}

where
\al{
  \mel\alpx\betx 
  = \Mel\alpx\betx - \mOmgo^{-1} \left[
    \vo\alpx \Mel\mPsi\betx
    -
    \lcsf\alpx\gamx \vo\gamx \NN\rPhi\betx
  \right] .
}
Thus, the net drift of a meandering spiral in an external field depends on the angle $\rPhi$ between the meander pattern and the applied field $\vE$. 

The average drift speed, measured over one meander period (\ie\ one petal of the flower) will be:
\alnn{ %eq{Vn}{}
  {\V^{\a}_{\nm}} =& {\dist^{\a}_{\nm}} / \To_\nm = \R\a\alpx(\rPhi_\nm) \Q\alpx\betx \avg{\vo\alpx}  \\ 
  & + \R\a\alpx(\rPhi_\nm)  \R\betx\b (\rPhi_\nm) \E^\b + \OO(\heta^2).
}
 
with 
\alnn{
  \Q\alpx\betx = \avg {\mel\alpx\betx} + \frac{\avg{\vo\alpx} \avg{\Mel\mPsi\betx}}{\mOmgo}
}

If we are outside the phase-locking regime, the ratio $\alp/2\pi = \romgo/\mOmgo$ will not be close to a fraction $\Mint/\Nint$ with $\Mint$ and $\Nint$ small integers. Therefore, after several rotations, the angle $\rPhi_\nm$ will have taken different values that are uniformly distributed over the unit circle. In that approximation, the time average of a quantity (net rotation, displacement or time lapse) will be that quantity averaged over all possible phases $\rPhi_\nm$ at the start of the meander cycle: 
\alnn{
  \mean{\To} & = \lim_{\mm\rightarrow \infty} \frac{\t_{\nm+\mm}-\t_\nm}{\mm} 
               = \frac{1}{2\pi} \int_0^{2\pi} \To_\nm(\rPhi_\nm) \,\d\rPhi_\nm ,   \\
  \mean{\alp} &= \lim_{\mm\rightarrow \infty} \frac{\rPhi_{\nm+\mm}-\rPhi_\nm}{\mm} 
               = \frac{1}{2\pi} \int_0^{2\pi} \alp_\nm(\rPhi_\nm) \,\d\rPhi_\nm ,  \\
  \mean{ \dist^\a } &= \lim_{\mm\rightarrow \infty} \frac{{\X^{\a}_{\nm+\mm}-\X^{\a}_{\nm}}}{\mm}
               = \frac{1}{2\pi} \int_0^{2\pi} {\dist^{\a}_{\nm}}(\rPhi_\nm) \,\d\rPhi_\nm ,  \\
  \mean{ \V^\a } &= \lim_{\mm\rightarrow \infty} \frac{{\X^{\a}_{\nm+\mm}-\X^{\a}_{\nm}}}{\t_{\nm+\mm} - \t_\nm} 
               = \frac{1}{2\pi} \int_0^{2\pi} \V^\a_\nm(\rPhi_\nm) \,\d\rPhi_\nm. 
}

Inserting Eqs. \eq{Tn}, \eq{alphan}, the average duration and rotation angle over a meander cycle satisfy
\alnn{
  \mean{\To} &=  \To + \OO(\heta^2), \\
  \mean{\alp} &= \alp + \OO(\heta^2). 
}

For the average drift speed, we then find
\al{
  \mean{\V^\a} = \Gone \E^\a + \Gtwo \lcsf\a\b \E^\b, \eqlabel{Vmean}  
}
where
\al{
  \Gone &= \frac{1}{2} \avg{\qel\alpx\alpx}, \eqlabel{G12}\\
  \Gtwo &= \frac{1}{2} \lcsf\alpx\betx \avg{\qel\betx\alpx} , \nn
}

Since the trace of a matrix is invariant and $\beps=\Mx{\lcsf\alpx\betx}$ 
is the generator
of rotations, we furthermore find that for any $2\times 2$ matrix
$\bQ$ and rotation matrix $\bR$ holds 
\alnn{ % eq{rotinv}{
  \Tr(\bR \bQ \bR^T) &= \Tr(\bR^T \bR \bQ) = \Tr(\bQ ),  \\
  \Tr(\beps \bR \bQ \bR^T) &= \Tr(\bR^T \beps \bR \bQ) =  \Tr(\bR^T \bR  \beps   \bQ)  = \Tr( \beps  \bQ)  .
}
 
Hence 
\alnn{
  \avg{ \qel\alpx\alpx} &= \avg{ \qel\A\A}, \\
  {\lcsf\alpx\alp \avg{\qel\alp\alpx} } &= { \lcsf\A\B \avg{\qel\B\A}} . 
}

Expression \eq{Vmean} was also computed in the center frame in~\cite{Dierckx:2017PRL} using a double Fourier series. We find the same result here using the discrete maps for ${\X^{\a}_{\nm}}$ and $\rPhi_\nm$. From Figs. \ref{fig:freeEFDBa} and \ref{fig:freeEFDFK} it can be seen that the net drift velocity is well described by Eq. \eq{Vmean} with coefficients \eq{G12}.

\section{Discussion \label{sec:discussion}}

In this work, we have introduced different coordinate systems that allow to describe the dynamics of one- and two-dimensional wave patterns. The origin of the chosen coordinate system is interpreted as the wave front or spiral wave tip position, or as the filament position for 3D scroll waves. 

For spiral waves, different coordinate systems produce the same result where they are applicable, but they can differ in their applicability areas. So, the ``minimally rotating'' frame of reference is not good for describing drift of meandering spirals for parameter values close to the equivariant Hopf bifurction (Winfree's $\partial M$ boundary), and one needs to use the ``co-rotating finite-core'' frame instead. Similarly, the ``center'' frame is inadequate for the parameters in the vicinity of the ``resonant meander'' parameters, and one is much better with the ``tip'' frame there.

From the simple application of electroforetic drift already, it can be seen that the time-scale at which the motion is analysed is important. Here, we have demonstrated that different time-averaging strategies can be taken in addition to the Fourier approach of \cite{Dierckx:2017PRL}: analysis of the return map and a sliding average are possible. 

We have deliberately included a brief description of the geometrical viewpoint on phase space dynamics in Sec. \ref{sec:geom}, which shows that our technical computations are part of a simpler geometric theory. 

We have given two applications of the response function framework here, both related to spiral wave drift in a constant external field. First, we have shown that the mean drift speed is equal in the center frame and finite-core frame, as both overlap integrals produce the same physical tension coefficients. Secondly, we have computed that for phase-locked meander, the drift speed is not proportional to the applied field strength, but close to the orbital velocity along the meander flower. 

% Tip frame calculations were not given here, but the here-developed framework can be used in future work to analyse and predict the dynamics of meandering scroll waves in a heterogenous medium such as cardiac tissue. 

As always, the limitations of the presented work suggest the directions for future research. We mentioned in the beginning of Section~\ref{sec:geom} that the mathematical foundation of our formal approach has known outstanding technical difficulties, \eg\ to describe spiral waves the reaction-diffusion system has to be considered in a Banach space in which the action of the Euclidean group is not even continuous, and that the effective spatial localization of the response functions is only an empirical fact. A possible route through the first obstacle was shown by~\cite{Sandstede:1997} which informally can be summarized thus: rather than considering the whole Banach space, one ought to focus on its part occupied by actual spiral-wave solutions, and in that part, the group acts ``nicely''. For the second difficulty, a promising rigorous result was obtained in~\cite{Sandstede-Scheel-2004}: that  response functions of one-dimensional analogues of spiral waves are indeed exponentially localized in space.

From the physical viewpoint, more interesting are the limitations related to the types of perturbations considered. Of special interest is wave propagation in confined irregular geometries such as cardiac tissue. Previous works have shown that these can also to some extent be described by a perturbative approach \cite{Biktasheva:2015, Marcotte:2015}, although the boundary perturbation may take the shape of a Dirac distribution and is therefore not small at all. 

Note that phase-locking of meandering spirals caused by perturbations that are purely time-dependent have been considered before~\cite{Grill:1996,Mantel-Barkley-1996}. Our approach outlined in~\cite{Dierckx:2017PRL} and detailed in here is potentially by far more universal. The simple examples considered so far are admittedly only first steps, and consideration of more realistic problems, such as cardiac tissue with its anisotropy and spatial heterogeneity, promises many new discoveries. 

This research was supported in part the EPSRC grants EP/E018548/1, EP/D074789/1, EP/P008690/1, EP/N014391/1 and EP/E016391/1 (UK) and National Science Foundation Grants No. NSF PHY-1748958, NIH Grant No. R25GM067110, and the Gordon and Betty Moore Foundation Grant No. 2919.01 (USA).

% \section*{References}
%\bibliography{references_HansVadim} 

%\bibliographystyle{unsrt} 

% \cleardoublepage

\appendix
\section*{Appendix: list of main notations}

Note that some symbols denote different objects in different places; we hope their meaning is sufficiently clear from the context.

\vspace*{1\baselineskip}

\begin{ruledtabular}
\begin{tabular}{@{$} l @{$} p{6.9cm}} % first column in math mode accents
\base{\ff} & unperturbed value of $\ff$ \\
\pert{\ff} & perturbed value of $\ff$ \\
\cen{\ff} & value of $\ff$ in the center frame  \\
\tip{\ff} & value of $\ff$ in the tip frame  \\
\avg{\ff} & average of $\ff$ over a cycle \\% (result independent of $\Psi$) 
\slavg{\ff} & sliding average of $\ff$ \\ % (result depends on $\Psi)$\\
% brackets
\braket{.}{.} & inner product in space \\
\bbraket{.}{.} & cycle-averaged inner product \\
\mean{.} & average over uniformly distributed initial conditions \\ 
% indices
\a,\b, ... & spatial indices in the lab frame of reference, in \{1,2\} \\
 \A,\B, ... & spatial indices in the moving (rotating) frame of reference \\
\apar, \bpar, \cpar & kinetic parameters of the Barkley model \\
\alpx, \betx & spatial indices in a frame that follows the unperturbed dynamics \\
\fhneps,\fhnbet,\fhngam & kinetic parameters of the FitzHugh-Nagumo model \\
 \alp & angle between petals of the meander flower \\
\c & 1D frame velocity \\
\c^\A & velocity components of the moving frame, expressed in the moving frame\\ 
\ccent & middle of unperturbed spiral wave tip trajectory \\
\Delta & Laplacian operator \\
\dist^\a& net displacement by the spiral wave tip over one meander period \\
\lcs & Levi-Civita symbol\\
%\meps & timescale ratio in the model \\ 
\shift^\a & components of infinitesimal translation vector \\
\vE & external field \\
\heta & order of magnitude of the external perturbation \\
  \bF & column-vector of $\Nv$ nonlinear functions representing reaction kinetics \\
\rPhi & rotation angle of the moving frame \\
\g & {element of transformation group} \\
\gone &scalar filament tension (circular core case) \\
\gtwo & pseudoscalar filament tension (circular core case) \\
\Group & system's symmetry group \\
\Gone &scalar filament tension (meander case) \\
\Gtwo & pseudo-scalar filament tension (meander case) \\
\Subgroup & solutions's spatiotemporal symmetry group \\
\hh & perturbation to the reaction-diffusion system\\
\j,\k & spatial indices in the lab frame of reference \\
\Jac & Jacobian matrix for the collective coordinates \\
\SSubgroup &solutions's spatial symmetry group \\
 \K & spiral wave chirality (-1 or +1) \\
\linop & linearized operator in the lab frame \\
\HL & linearized operator in the moving frame \\
\m,\n, ... & generalized indices in the lab frame of reference, in $\{\x, \y, \rPhi, \mPsi \}$ \\
\M, \N, ...& generalized indices in the moving frame, in $\{1, 2, \rPhi, \mPsi \} $\\
\MM & coupling matrix\\
\Nv & number of state variables in the reaction-diffusion system\\
 \Ns & number of spatial dimensions\\
\NBS & number of broken continuous spatiotemporal symmetries \\
{\NBSS\quad } & number of broken continuous spatial symmetries \\
\NBTS & number of broken continuous temporal symmetries (0 or 1) \\
\orig & origin of the lab frame \\
 \romgo & instantaneous rotation velocity of the spiral wave  \\
\avg{\romgo} & cycle-averaged rotation velocity of the spiral wave \\
  \mOmgo & frequency of temporal cycle completion ($\mOmgo = 2\pi /\To$) \\
\HP& matrix of diffusivities\\
\mPsi & temporal phase of the relatively periodic pattern / meander phase \\
\Quot & quotient space \\
\r &position vector in the lab frame \\
\R\a\A & rotation matrix \\
\xp & spatial coordinate in the moving frame of a point ($\xp = \x -\X$) \\
\t & time in the lab frame  \\
\tp  & time in the moving frame\\
\Trot & time needed to describe a full meander flower \\ 
\To & duration of a meander cycle / relative orbit \\
\rphi & polar angle in the moving frame $(\rphi = \polang - \rPhi)$\\
\polang & polar angle in the lab frame\\
\uu & column-vector of state variables \\
\u_\j & $\j$-th state variable \\
\uc\j & pinning constant \\
\uuo & unperturbed spiral solution in the moving frame \\
\UU & solution to the reaction-diffusion equation in the lab frame (column-vector) \\
\VV_\M & critical eigenmode \\
\PS & phase space \\
\pv\M & instantaneous drift speed \\
\V^\a & components of average spiral wave drift speed\\
\ww^\m & response function / critical adjoint eigenmode in lab frame \\
\WW^\M & response function / critical adjoint eigenmode in a moving frame \\
\x^\a & lab frame coordinates of a point \\
\x,\y & lab frame coordinates of a point \\
\X,\; \X^\a & lab frame coordinate(s) of the origin of the moving frame \\
 \pfil & lab frame coordinate of the origin of the moving frame \\
\X^\M & \mbox{} collective coordinates \\
\cfil & origin of the moving frame \\
\Claytonphase & polar angle in kinetics phase space \\
 \ctip & wave front / spiral wave tip position\\
 \ptip & wave front / spiral wave tip coordinate in the lab frame\\
 \dfil & wave front / spiral wave tip coordinate in the moving frame
\end{tabular}
\end{ruledtabular}

\end{document}